\documentclass[journal]{IEEEtran} 

\usepackage[utf8]{inputenc}
\usepackage{amsmath,amssymb,amsfonts,amsthm}
\usepackage[numbers,sort&compress]{natbib}
\usepackage{bm}
\usepackage{color}
\usepackage{glossaries}
\usepackage{graphicx}
\usepackage[caption = false]{subfig}
\usepackage{url}
\usepackage{enumitem}
\newacronym{BS}{BS}{base station}
\newacronym{DMA}{DMA}{dynamic metasurface antenna}
\newacronym{FAMA}{FAMA}{fluid antenna multiple access}
\newacronym{FAS}{FAS}{fluid antenna system}
\newacronym{MIMO}{MIMO}{multiple-input multiple-output}
\newacronym{PCB}{PCB}{printed circuit board}
\newacronym{PDF}{PDF}{probability density function}
\newacronym{PEC}{PEC}{perfect electric conductor}
\newacronym{RF}{RF}{radio-frequency}
\newacronym{SIR}{SIR}{signal-to-interference ratio}
\newacronym{SIW}{SIW}{substrate integrated waveguide}

\renewcommand{\vec}[1]{\mathbf{\lowercase{#1}}}	   
\newcommand{\mat}[1]{\mathbf{\uppercase{#1}}}	   
\newcommand{\real}[1]{\text{Re}\left\{#1\right\}}	
\newcommand{\imag}[1]{\text{Im}\left\{#1\right\}}	
\newtheorem{remark}{Remark}

\title{Metasurface-based Fluid Antennas: from Electromagnetics to Communications Model}

\author{Pablo Ramírez-Espinosa,~\IEEEmembership{Member,~IEEE}, Cleofás Segura-Gómez, Ángel Palomares-Caballero, F. Javier López-Martínez,~\IEEEmembership{Senior Member,~IEEE}, and David Morales-Jiménez,~\IEEEmembership{Senior Member,~IEEE}.
\thanks{The work by P. Ram\'irez-Espinosa has been funded by the European Union under the Marie Sklodowska-Curie grant agreement No. 101109529. This work is also supported by grant PID2023-149975OB-I00 (COSTUME), PDC2023-145862-I00 and FPU20/00256 funded by MICIU/AEI/10.13039/501100011033 and FEDER/UE.}
\thanks{P. Ram\'irez-Espinosa is with the Telecommunications Research Institute (TELMA), University of M\'alaga,  29071 M\'alaga, Spain. E-mail: pre@ic.uma.es.}
\thanks{C. Segura-Gómez, Á. Palomares-Caballero, F. J. López-Martínez and D. Morales Jiménez are with the Department of Signal Theory, Networking and Communications, Research Centre for Information and Communication Technologies (CITIC-UGR), University of Granada, 18071, Granada, Spain. E-mails: \{cleofas, angelpc, fjlm, dmorales\}@ugr.es.}
\thanks{This work has been submitted to the IEEE for publication. Copyright may be transferred without notice, after which this version may no longer be accesible.}
}

\begin{document}
\maketitle

\begin{abstract}
    Fluid antenna systems (FASs) have become a popular topic in the wireless community as an effective yet simple means of exploiting spatial diversity. Due to the limitations of physically moving radiating elements, electronically reconfigurable antennas are emerging as practical implementations of FASs, since changing the radiation pattern is functionally equivalent to physically moving the device. However, electronically reconfigurable antennas pose a challenge in terms of analytical modeling, often requiring full-wave simulations or measurements for their characterization; this severely limits the extraction of theoretical insights useful for system design. Motivated by these difficulties and the growing interest in FASs, we propose in this paper a complete analytical model for metasurface-based embodiments of FASs. Specifically, we advocate for the implementation of the FAS concept through dynamic metasurface antennas (DMAs), hitherto proposed as array replacements in multiple-input multiple-output (MIMO) systems. 
    We leverage circuit theory to rewrite the conventional signal model of FASs in terms of admittance matrices accounting for the electromagnetic effects inherent to metasurfaces. The model is validated with full-wave simulations, showing good agreement. We further illustrate how to apply the model for standard performance analysis, and provide closed-form expressions for key metrics, including the resulting signal covariance matrix. Results confirm that practical DMA-based FASs can achieve similar performance to that of idealized implementations of position-flexible antennas. 
\end{abstract}

\begin{IEEEkeywords}
    Circuit modeling, dynamic metasurface antenna (DMA), electronically reconfigurable antenna, fluid antenna system (FAS), multiple access, spatial correlation. 
\end{IEEEkeywords}

%------------------------------------------------------------------------------------------------------------------
% Introduction
%------------------------------------------------------------------------------------------------------------------
\section{Introduction}
\IEEEPARstart{R}{ecently}, \glspl{FAS} have attracted considerable attention from the wireless community as a simple, practical and efficient extension of classical spatial diversity systems based on multiple fixed antennas. Conceptually, \gls{FAS} refers to a single antenna element---fed hence by a single \gls{RF} chain---that can be freely moved within a predefined aperture \cite{Wong2022_BruceLee}. Depending on the application, the antenna position that maximizes the \gls{SIR} or any other metric of interest is selected out of a dense grid of possible positions (ports), ideally \textit{sampling} the spatial channel at any point in the space continuum. Compared to conventional diversity systems with antenna spacing (sampling) at half-wavelength, \gls{FAS}s `oversample' the channel by, in principle, allowing ports to be arbitrarily close to one another. Thus, the opportunistic port selection provides a means to fully exploit the channel's spatial degrees of freedom, resulting in very substantial performance gains.    

Among the potential use cases of \glspl{FAS}, \gls{FAMA} is the flagship application, allowing to simultaneously serve several users without the need of precoding at the \gls{BS} or of interference cancellation \cite{Wong2022}. Thus, \gls{FAMA} arises as a promising solution to increase the multiplexing capabilities of a system without incurring in an excessive complexity or power consumption. Naturally, these capabilities are dominated by the spatial correlation, as close-by spatial locations have inherently correlated channels, limiting the achievable spatial degrees of freedom of the channel and, ultimately, the performance. Theoretical studies, where the correlation structure is approximated by different models, predict that a moderately high number of users can be simultaneously served, being this number larger as the physical aperture of the antenna increases \cite{Ramirez2024, Zhu2024, New2024, Khammassi2023}. Early studies on a fast-switching version of \gls{FAMA} predict massive multiple access capabilities (dozens of users simultaneously served), but this is thus far deemed unfeasible due to implementation limitations.

Indeed, despite the potential of \glspl{FAS}, with simple and energy-efficient solutions that offer very promising multiplexing capabilities, critical challenges arise in terms of practical physical implementation, i.e., as to how to realize the \glspl{FAS} concept in practice. Different alternatives have been proposed in the literature, including movable antennas \cite{Zhu2024_movable, Zhu2024_magazine, Dong2024}, liquid antennas \cite{Naqvi2019, Huang2021, Wang2025_arxiv}, or surface-wave based antennas \cite{Shen2024}. In these types of implementations, either the antenna itself is physically moved using micro-motors, or some liquid or metal is moved within the antenna by using pumps to achieve fine spatial sampling or to control the radiation properties. However, they all resort on physical movement to modify the antenna position or characteristics and, ultimately, to sample the spatial channel at different positions. Switching speed is critically important to realize the promising (massive) multiplexing capabilities of \gls{FAMA}, e.g., in fast-\gls{FAMA}, antenna positions need to be switched at the transmitted symbol rate \cite{Ramirez2024}. However, these mechanical devices are relatively slow, prone to errors, difficult to implement, and the energy required to operate motors or pumps may jeopardize the overall benefits of the energy efficiency brought by \gls{FAMA}. Aware of these limitations, the focus has recently shifted to electronically reconfigurable implementations, where any mechanical movement is avoided. To the best of our knowledge, two solutions have been proposed thus far along this line: a metasurface-based implementation in \cite{Liu2025}, and a pixel-based antenna in \cite{Zhang2024}. The former deploys a matrix of meta-elements on top of a radiating structure, which is attached to an \gls{RF} chain, selecting the corresponding element through a network of p-i-n diodes and hence \textit{virtually} moving the antenna. This approach may however suffer from low efficiency due to the small size of the meta-element and the fact that only one is active at a time. In turn, \cite{Zhang2024} proposes tuning the interconnection of several pixels (patch antennas) so that the radiation pattern of the whole structure changes. This effectively \textit{emulates} a \gls{FAS}, since changing the radiation pattern can indeed be equivalent to sampling the spatial channel at different positions (ports). Of course, closer configurations of the pixel antenna (similar radiation patterns) lead to higher spatial correlation, as in the case of adjacent ports in the original (idealized) \gls{FAS} concept.

Electronically reconfigurable \glspl{FAS} suggest that, in contrast to the theoretical models in \cite{Wong2022, Khammassi2023, Ramirez2024}, the performance limit is not only dictated by the spatial correlation process itself, but also by the ability of the device to change its radiation properties; and hence, the way the different impinging waves are combined. A follow-up question is, then, how to design these devices and their different configurations (radiation patterns) to maximize the delivered performance; and whether the performance barrier, dictated by the widely-adopted Jakes' and Clarke's correlation models, can be surpassed by suitably changing the effective spatial correlation (via suitable design of the configurations). Unfortunately, addressing these questions is by no means trivial, since electronically reconfigurable antennas as those in \cite{Liu2025,Zhang2024} lack a generalized signal model and one needs to rely on full-wave simulations or measurements to characterize them. Circuit theory can be combined with simulations to partially alleviate the analysis, as in \cite{Zhang2024}, but this modeling approach is inconveniently inflexible and unable to accommodate (even minor) adjustments in the antenna design, such as the dielectric permittivity, spacing between elements, or their interconnections.  

Aiming to provide the necessary tools to tackle the previous research questions, and motivated by the growing popularity of \glspl{FAS}, we propose in this paper a general yet tractable analytical framework to characterize \gls{DMA}-based \glspl{FAS}. Recently, \glspl{DMA} have become popular in the wireless literature as replacements for fully-digital and hybrid \gls{MIMO} arrays, promising reduced hardware cost and consumption without compromising the performance \cite{Schlezinger2021, Carlson2024, Williams2022, You2023, heath2025tri, Ramirez2025}. \glspl{DMA} admit a simple implementation by stacking several waveguide-fed metasurface antennas as in \cite{boyarsky2021_sciReports}, where the radiating elements (usually called meta-elements) are reconfigured through semiconductor devices \cite{Sleasman2016,Yoo2023_AWPL}. Changing the configuration of a \gls{DMA} (i.e., states of the semiconductor devices) implies that its radiation pattern is modified \cite{Li2021_TAP,Jabbar2024}. Therefore, simplified versions of \glspl{DMA} (simpler than the ones used for \gls{MIMO}) can be used as fluid antennas \cite{Liu2025}, being more scalable and simple than, e.g., pixel-based antennas \cite{Zhang2024}. For instance, \glspl{DMA} readily admit planar topologies with arbitrary sizes, whereas in pixel antennas, one needs to ensure an efficient illumination of the upper pixel layer, resulting in less compact designs.

In this paper, building upon circuit theory, we leverage and extend the modeling in \cite{Williams2022} to incorporate all relevant electromagnetic phenomena inherent to metasurfaces into the conventional \gls{FAS} signal model. This allows the characterization of practical effects such as insertion losses, reflections, and radiation properties in a straightforward manner. Our goal is to provide a general framework for analyzing the performance limits of \glspl{FAS} based on physically attainable and realistic implementations, without compromising the mathematical tractability. Specifically, our contributions are:
\begin{itemize}
    \item We generalize the \gls{DMA} circuital model in \cite{Williams2022} to account for arbitrary loads at the waveguides (allowing, e.g., to connect two \gls{RF} chains to each waveguide), and particularize it for a fluid antenna implementation based on a \gls{DMA} controlled by p-i-n diodes.
    \item The fluid antenna implementation is validated through full-wave simulations, showing good agreement. 
    \item The circuital model is mapped to the conventional \gls{FAS} communication model, yielding expressions in the same form as the theoretical (and idealized) signal model in the \gls{FAS}-related literature (e.g., \cite{Khammassi2023, Wong2022, Wong2022_EL, Ramirez2024}), but accounting now for the electromagnetic phenomena inherent to the \gls{DMA} structure. 
    \item We provide a closed-form expression for the resulting covariance matrix of the effective baseband channel (spatial correlation) in terms of: \textit{i)} the wireless channel covariance (imposed by the environment) and \textit{ii)} the \gls{DMA} topology (controllable by design and \mbox{p-i-n} diode configuration). 
    \item We exemplify the utility of the model through \gls{FAMA} performance evaluation using different antenna configurations, showing that a generic \gls{DMA}-based \gls{FAS} can match the performance delivered by a conceptual reference \gls{FAS} where the antenna element can be freely moved. 
\end{itemize}

The remainder of this paper is organized as follows: Section \ref{sec:DMA_Model_General} introduces the circuit modeling of generic \glspl{DMA}. Section \ref{sec:FAS_DMA_Model} specializes the model for a \mbox{p-i-n} diode-controlled \gls{DMA} \gls{FAS}, provides closed-form expressions for relevant metrics such as the radiated power, and validates the design through full-wave simulations. Section \ref{sec:FAS_Model} maps the circuit model into the conventional \gls{FAS} signal model, and analyzes the resulting correlation structure of the effective channel. Section \ref{sec:Performance} evaluates the performance of \gls{FAMA} when using the proposed \gls{FAS} design. Finally, Section \ref{sec:Conclusions} presents the key conclusions and outlines future research directions.  

\textit{Notation:} Vectors and matrices are represented by bold lowercase and uppercase symbols, respectively. $(\cdot)^T$ and  $(\cdot)^H$ denote the matrix transpose and the conjugate transpose, and $\|\cdot\|_2$ is the $\ell_2$ norm of a vector. $(\mat{A})_{j,k}$ is the $j,k$-th element of $\mat{A}$, and $\mat{I}_N$ is the identity matrix of size $N\times N$. $\mathbb{E}[\cdot]$, $\real{\cdot}$, and $\imag{\cdot}$ are the mathematical expectation and the real and imaginary parts operators. Finally, $i = \sqrt{-1}$ is the imaginary number. Any other specific notation will be defined when necessary.

\textit{Supplementary files:} The MATLAB code used for simulations and figures in this paper will be available on Github upon publication of the manuscript.

%can be found at https://github.com/preugr/Fluid-antenna-block-diagonal-model.git.

%------------------------------------------------------------------------------------------------------------------
% CIRCUITAL MODEL OF DMAs
%------------------------------------------------------------------------------------------------------------------
\section{General Circuital Model of DMAs}
\label{sec:DMA_Model_General}
\begin{figure}[t]
    \centering
    \includegraphics[width=1\linewidth]{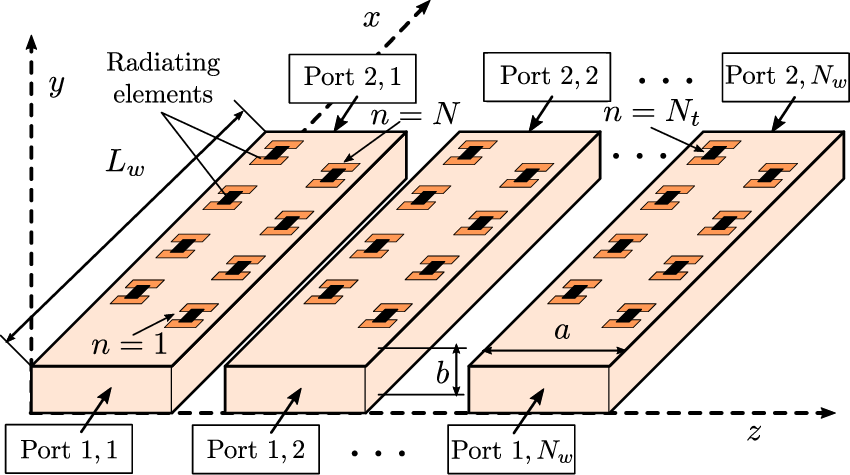}
    \caption{Schematic of DMA with multiple stacked waveguides.}
    \label{fig:DMA_schematic}
\end{figure}

Consider a generic \gls{DMA} consisting of $N_w$ identical rectangular waveguides, as illustrated in Fig. \ref{fig:DMA_schematic}, which are terminated in two ports (one at each end) representing either an \gls{RF} chain or a load. The waveguides are placed on the $xz$-plane, with dimensions: $a$ (height) along the $z$-axis, $b$ (width) along the $y$-axis, and $L_w$ (length) along the $x$-axis, being $a$ and $b$ chosen such that only the fundamental mode $\text{TE}_{10}$ propagates. For generality, we assume the waveguides are filled with a lossless dielectric material of permittivity $\varepsilon$. 

On the upper face ($y = b$) of each waveguide, $N$ radiating elements are attached at arbitrary positions, with resonance frequencies controlled by some semiconductor device (e.g., varactors or p-i-n diodes), making the concept of \gls{DMA} equivalent to stacking multiple waveguide-fed metasurface antennas or reconfigurable leaky-wave antennas \cite{Smith2017}. The radiating elements are modeled as sub-wavelength magnetic dipoles, with such modeling assumption validated in, e.g., \cite{Smith2017, Williams2022, Johnson2014, Jabbar2024}. The \gls{DMA} is used to communicate with $M$ single antenna devices, which are also modeled as magnetic dipoles without loss of generality. These devices can represent users in a cellular network or different \gls{BS} antennas according to the general \gls{FAMA} communication model (see, e.g., \cite{Wong2022}). 

Next, we analyze the proposed system when using the \gls{DMA} as transmitter and as receiver. To that end, we inherit the circuital model in \cite{Williams2022}, and generalize it to account for the additional ports in the waveguides\footnote{Note that the model in \cite{Williams2022} assumes the waveguides are terminated in a metallic wall, and we here relax this assumption by allowing an arbitrary load at every waveguide port.}.

%----------------------------------------------
% DMA as TRANSMIT ANTENNA (DOWNLINK)
%----------------------------------------------
\subsection{DMA as transmitter}

\begin{figure}[t]
    \centering
    \includegraphics[width=1\linewidth]{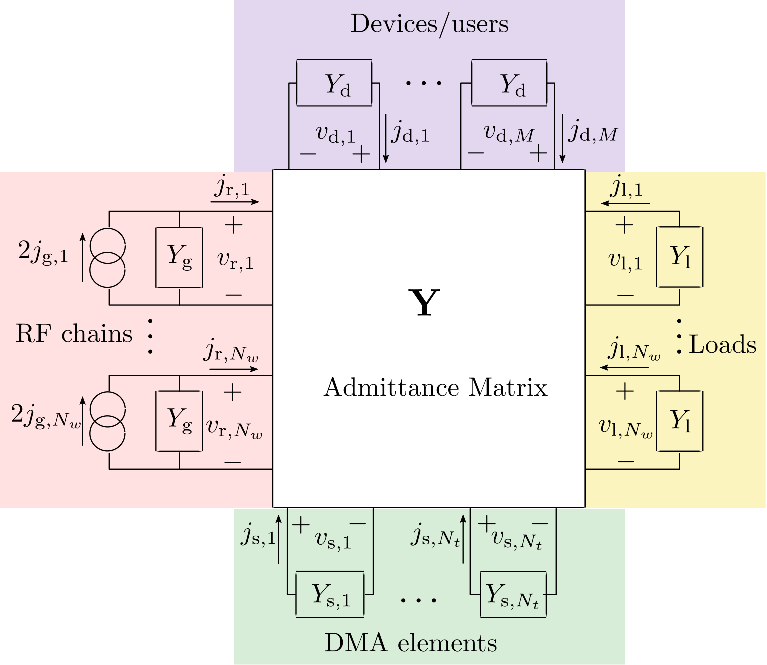}
    \caption{Multiport representation of DMA-based communications system for transmission.}
    \label{fig:DMA_multiport}
\end{figure}

We consider, without loss of generality, that the waveguides are fed at Ports 1 ($x = 0$ according to Fig. \ref{fig:DMA_schematic}), while Ports 2 ($x = L_w$) are generic loads. The fed signals propagate through the waveguides, exciting the radiating elements which scatter part of the signal power out. Since the elements are modeled as magnetic dipoles, we formulate a circuital multiport network in terms of magnetic voltages and currents, which is completely described by the mutual admittances. Thus, extending the model in \cite{Williams2022}, the system is described by the multiport network in Fig. \ref{fig:DMA_multiport}, i.e.,
\begin{equation}
    \begin{bmatrix}\vec{v}_\text{r} \\ \vec{v}_\text{s} \\ \vec{v}_\text{l} \\
    \vec{v}_\text{d}\end{bmatrix} = \begin{bmatrix}\mat{Y}_\text{rr} & \mat{Y}_\text{sr}^T & \mat{Y}_\text{lr}^T & \mat{0} \\
         \mat{Y}_\text{sr} & \mat{Y}_\text{ss} & \mat{Y}_\text{sl} & \mat{Y}_\text{ds}^T \\
         \mat{Y}_\text{lr} & \mat{Y}_\text{sl}^T & \mat{Y}_\text{ll} & \mat{0}\\
         \mat{0} & \mat{Y}_\text{ds} & \mat{0} & \mat{Y}_\text{dd}\end{bmatrix}\begin{bmatrix}
             \vec{j}_\text{r} \\ \vec{j}_\text{s} \\      \vec{j}_\text{l}\\
             \vec{j}_\text{d}
         \end{bmatrix}, \label{eq:CircuitModelTx}
\end{equation}
where $\vec{v}_\text{r}, \vec{j}_\text{r}\in\mathbb{C}^{N_w\times 1}$ are the magnetic voltages (measured in Amperes) and currents (measured in Volts) at the \gls{RF} chains interface (entering the waveguides from Ports 1), $\vec{v}_\text{s}, \vec{j}_\text{s}\in\mathbb{C}^{N_t\times 1}$ with $N_t = N\times N_w$ are the voltages and currents at the radiating elements, and $\vec{v}_\text{l}, \vec{j}_\text{l}\in\mathbb{C}^{N_w\times 1}$ and $\vec{v}_\text{d}, \vec{j}_\text{d}\in\mathbb{C}^{M\times 1}$ are the corresponding voltages and currents at the waveguides loads (Ports 2) and single antenna devices, respectively. Besides, $\mat{Y}_\text{mn}$ is the mutual admittance matrix between agent $\text{n}$ and $\text{m}$, for which closed-form expressions will be provided later. Importantly, $\mat{Y}_\text{ds}$ is the mutual admittance between radiating elements and devices, hence representing the wireless propagation channel. 

As illustrated in Fig. \ref{fig:DMA_multiport}, each \gls{RF} chain (Port 1) is modeled as an equivalent current generator with intrinsic admittance $Y_\text{g}\in\mathbb{C}$ and available magnetic current $j_{\text{g},n}\in\mathbb{C}$ with $n = 1,\dots, N_w$, while each Port 2 has a load $Y_\text{l}\in\mathbb{C}$ and the single antenna devices are loaded with an admittance $Y_\text{d}\in\mathbb{C}$ (for simplicity, the same for all devices). The reconfigurability of the radiating elements is modeled by attaching to them a tunable load admittance $Y_{\text{s},n}\in\mathbb{C}$ for $n = 1,\dots, N_t$. For notation simplicity, we define the diagonal matrix $\mat{Y}_\text{s}\in\mathbb{C}^{N_t\times N_t}$ such that $(\mat{Y}_\text{s})_{n,n} = Y_{\text{s},n}$. 

The objective is to characterize the induced currents $\vec{j}_\text{d}$ at the devices in terms of the supplied currents $\vec{j}_\text{g} = \setlength\arraycolsep{3pt}\begin{pmatrix}
    j_{\text{g},1} & \dots & j_{\text{g},N_w}
\end{pmatrix}^T$. Considering the system operates out of the reactive near field, we can neglect backscattering\footnote{This, known as \textit{unilateral assumption}, considers that pathloss is large enough so that the currents $\vec{j}_\text{d}$ do not couple to the transmitting antennas.} and, hence, assume $\mat{Y}_\text{ds}^T = \mat{0}$ in \eqref{eq:CircuitModelTx}. Under this assumption, from \eqref{eq:CircuitModelTx} and using $\vec{v}_\text{l} = -Y_\text{l}\vec{j}_\text{l}$ and $\vec{v}_\text{s} = -\mat{Y}_\text{s}\vec{j}_\text{s}$ (Ohm's law), we obtain
\begin{equation}
    \vec{j}_\text{d} = (Y_\text{d}\mat{I}_M + \mat{Y}_\text{dd})^{-1}\mat{Y}_\text{ds}\left(\mat{Y}_\text{s} + \widetilde{\mat{Y}}_\text{ss}\right)^{-1}\widetilde{\mat{Y}}_\text{sr}\vec{j}_\text{r}, \label{eq:jd_tx}
\end{equation}
where, for the sake of notation, we have defined the matrices
\begin{align}
    \widetilde{\mat{Y}}_\text{ss} &= \mat{Y}_\text{ss} - \mat{Y}_\text{sl}(Y_\text{l}\mat{I}_{N_w} + \mat{Y}_\text{ll})^{-1}\mat{Y}_\text{sl}^T, \\
    \widetilde{\mat{Y}}_\text{sr} &= \mat{Y}_\text{sr} - \mat{Y}_\text{sl}(Y_\text{l}\mat{I}_{N_w} + \mat{Y}_\text{ll})^{-1}\mat{Y}_\text{lr}.
\end{align}
Note that if $Y_\text{l}\rightarrow\infty$ (which implies that the waveguides are terminated in a metallic wall), then $\widetilde{\mat{Y}}_\text{ss} \rightarrow \mat{Y}_\text{ss}$ and $\widetilde{\mat{Y}}_\text{sr} \rightarrow \mat{Y}_\text{sr}$, boiling down to the model in \cite{Williams2022}. On the other hand, the currents entering the waveguides are related to the supplied ones through
\begin{equation}
    \vec{j}_\text{r} = 2Y_\text{g}(Y_\text{g}\mat{I}_{N_w}+\mat{Y}_\text{p})^{-1}\vec{j}_\text{g}, \label{eq:Rel_jr_jg}
\end{equation}
where $\mat{Y}_\text{p}$ is defined as $\vec{v}_\text{r} = \mat{Y}_\text{p}\vec{j}_\text{r}$, i.e.,
\begin{align}
    \mat{Y}_\text{p} =& \mat{Y}_\text{rr} - \widetilde{\mat{Y}}_\text{sr}^T(\mat{Y}_\text{s}+\widetilde{\mat{Y}}_\text{ss})^{-1}\widetilde{\mat{Y}}_\text{sr} \notag \\
    &-\mat{Y}_\text{lr}^T(Y_\text{l}\mat{I}_{N_w} + \mat{Y}_\text{ll})^{-1}\mat{Y}_\text{lr}. \label{eq:Yp}
\end{align}
Notice that $\mat{Y}_\text{p}$ can be seen as the input admittance matrix of the waveguides. In fact, if $\mat{Y}_\text{p} = Y_\text{g}\mat{I}_{N_w}$, then $\vec{j}_\text{r} = \vec{j}_\text{g}$ and all the available power is introduced into the waveguides. Then, as expected, it is a function of the configuration of the radiating elements $\mat{Y}_\text{s}$, their position in the waveguides (through the mutual coupling $\widetilde{\mat{Y}}_\text{ss}$) and the waveguide load $Y_\text{l}$. 

From \eqref{eq:jd_tx}, \eqref{eq:Rel_jr_jg} and \eqref{eq:Yp}, the end-to-end communication system is completely described in terms of the admittance parameters. 

%----------------------------------------------
% DMA as RECEIVING ANTENNA
%----------------------------------------------
\subsection{DMA as receiver}

In this case, the circuital model remains the same, i.e., the system is again described by \eqref{eq:CircuitModelTx}. The difference now is that the single antenna devices transmit the desired signal---i.e., $\vec{j}_\text{d}$ is the independent vector---and the \gls{DMA} acts as receiver. Thus, each port at the waveguides is modeled as a load, representing either an attached receiver or a waveguide termination. Referring to Fig. \ref{fig:DMA_multiport}, now the current generators are placed at the devices.

As the communications direction is also reversed, $\mat{Y}_\text{ds}$ represents now the backscattering channel, and hence can be neglected ($\mat{Y}_\text{ds}=\mat{0}$). Assuming this, the received signals at both ports in the waveguides are obtained from \eqref{eq:CircuitModelTx} and Ohm's law as
\begin{align}
    \vec{j}_\text{r} &= (Y_\text{g}\mat{I}_{N_w} + \mat{Y}_\text{p})^{-1}\widetilde{\mat{Y}}_\text{sr}^T(\mat{Y}_\text{s} + \widetilde{\mat{Y}}_\text{ss})^{-1}\mat{Y}_\text{ds}^T\vec{j}_\text{d}, \label{eq:jr_rx} \\
     \vec{j}_\text{l} &= (Y_\text{l}\mat{I}_{N_w} + \mat{Y}_\text{q})^{-1}\widetilde{\mat{Y}}_\text{sl}^T(\mat{Y}_\text{s} + \widehat{\mat{Y}}_\text{ss})^{-1}\mat{Y}_\text{ds}^T\vec{j}_\text{d}, \label{eq:jl_rx}
\end{align}
where we have kept the notation $Y_\text{g}$ and $Y_\text{l}$ for coherence with the previous case, and we have defined the matrices $\mat{Y}_\text{q}$, $\widehat{\mat{Y}}_\text{ss}$ and $\widetilde{\mat{Y}}_\text{sl}$ as the counterparts of $\mat{Y}_\text{p}$, $\widetilde{\mat{Y}}_\text{ss}$ and $\widetilde{\mat{Y}}_\text{sr}$, i.e.,
\begin{align}
        \mat{Y}_\text{q} &= \mat{Y}_\text{ll} - \widetilde{\mat{Y}}_\text{sl}^T(\mat{Y}_\text{s}+\widehat{\mat{Y}}_\text{ss})^{-1}\widetilde{\mat{Y}}_\text{sl} \notag \\
        &\quad\quad- \mat{Y}_\text{lr}(Y_\text{g}\mat{I}_{N_w} + \mat{Y}_\text{rr})^{-1}\mat{Y}_\text{lr}^T, \label{eq:Yq} \\
        \widehat{\mat{Y}}_\text{ss} &= \mat{Y}_\text{ss} - \mat{Y}_\text{sr}(Y_\text{g}\mat{I}_{N_w} + \mat{Y}_\text{rr})^{-1}\mat{Y}_\text{sr}^T, \\
        \widetilde{\mat{Y}}_\text{sl} &= \mat{Y}_\text{sl} - \mat{Y}_\text{sr}(Y_\text{g}\mat{I}_{N_w} + \mat{Y}_\text{rr})^{-1}\mat{Y}_\text{lr}^T.
\end{align}

Naturally, the system would only have access to $\vec{j}_\text{r}$ and $\vec{j}_\text{l}$ if a receiver is attached at the corresponding ports (Port 1 for $\vec{j}_\text{r}$ and Port 2 for $\vec{j}_\text{l}$). Note also that both \eqref{eq:jr_rx} and \eqref{eq:jl_rx} are the reciprocal of \eqref{eq:jd_tx}, as expected.

To complete the input-output relationship, we relate $\vec{j}_\text{d}$ with the supplied currents\footnote{With a slight abuse of notation, the supplied current is also denoted by $\vec{j}_\text{g}$, as in the case of transmission mode (where the current was supplied by the \gls{RF} chain). This allows for better clarity in the presentation.} $\vec{j}_\text{g}$ (now supplied by the single antenna devices). Similar to the transmit case, we now have
\begin{equation}
    \vec{j}_\text{d} = 2Y_\text{d}(Y_\text{d}\mat{I}_M + \mat{Y}_\text{dd})^{-1}\vec{j}_\text{g}. \label{eq:Rel_jd_jg}
\end{equation}

%----------------------------------------------
% ADMITTANCES EXPRESSIONS
%----------------------------------------------
\subsection{Expressions for mutual admittances}

Every agent in the system (radiating element, waveguide port, or user) is represented as a port in the network in \eqref{eq:CircuitModelTx}. Therefore, the mutual admittance $Y_{m,n}$ defines the interaction between port $n$ (source) and $m$ (destination). For the calculation of the admittances, we denote the position of each \gls{DMA} element by 
$\vec{r}_n=\setlength\arraycolsep{3pt}\begin{pmatrix}x_n & y_n & z_n\end{pmatrix}^T$ for $n= 1,\dots,N_t$, with $y_n = b$ according to Fig. \ref{fig:DMA_schematic}. Similarly, each single antenna device is located at arbitrary coordinates  $\vec{r}_m=\setlength\arraycolsep{3pt}\begin{pmatrix}x_m & y_m & z_m\end{pmatrix}^T$ for $m = 1,\dots,M$. As mentioned before, all these antennas are modeled as magnetic dipoles of infinitesimal length $l_m\in\mathbb{R}^+$ oriented along the $z$-axis, which implies that they only react to the $z$-component of the impinging magnetic field, and that the electrical Green's functions can be assumed constant along the dipole. The dipole length $l_m$ is a mere constant, scaling the whole system, and hence can be set to $1$ without compromising the validity of the model. However, for mathematical rigor, we explicitly state it in the admittance expressions. For convenience, we consider that the waveguide ports (both Ports 1 and 2) are also magnetic dipoles with the same orientation. Besides, we define $\omega$ as the angular frequency, $k_0$ and $k$ as the wavenumbers in the air and the waveguide dielectric, $\varepsilon_0$ as the permittivity of the air, and $k_x = \sqrt{k^2-\pi^2/a^2}$ as the propagation constant for the mode $\text{TE}_{10}$.

To obtain closed-form expressions for each matrix, we follow the general procedure described in \cite{Williams2022}: \textit{i)} port $n$ (source) is excited with an impulsive magnetic current, while the rest of ports are short-circuited so that there is no magnetic current flowing through them, \textit{ii)} the induced magnetic voltage at port $m$ (destination) is computed as the integral of the $z$-component of the impinging field, \textit{iii)} the mutual admittance is the ratio of the induced voltage and current, yielding \cite[Eq. (31)]{Williams2022}
\begin{equation}
    Y_{m,n} = il_m^2\omega\varepsilon G_{zz}(\vec{r}_m,\vec{r}_n), \label{eq:GeneralMutualAdmittance}
\end{equation}
with $G_{zz}(\cdot)$ the $z$-to-$z$ component of the corresponding dyadic Green's function, and $\varepsilon$ the corresponding permittivity. Next, we particularize \eqref{eq:GeneralMutualAdmittance} for each mutual admittance in \eqref{eq:CircuitModelTx}.

\subsubsection{$\mat{Y}_\textnormal{sr}$ and $\mat{Y}_\textnormal{sl}$} these matrices represent the coupling from the waveguides Ports 1 and 2 to the radiating elements, respectively---that is, what arrives to each radiating element from the signal introduced at each port. When the elements are short-circuited (as required to calculate the admittance), they are electrically invisible and hence they do not interact with the field \cite{Rogers1986}. This implies that the waveguides behave as if no elements were attached to them. Then, we straightforwardly have that $(\mat{Y}_\textnormal{sr})_{n,p} = (\mat{Y}_\textnormal{sl})_{n,p} = 0$ if element $n$ is not in the same waveguide than port $p$, and otherwise the mutual admittance is given by 
\begin{align}
    (\mat{Y}_\textnormal{sr})_{n,p} &= il_m^2\omega\varepsilon G_w\left(\overline{\vec{r}}_n,\overline{\vec{r}}_{\text{p}1}\right), \\
      (\mat{Y}_\textnormal{sl})_{n,p} &= il_m^2\omega\varepsilon G_w\left(\overline{\vec{r}}_n,\overline{\vec{r}}_{\text{p}2}\right)
\end{align}
for $n = 1,\dots,N_t$, where $\overline{\vec{r}}_{\text{p}1} = \setlength\arraycolsep{3pt}\begin{pmatrix}0 & y_p & z_p\end{pmatrix}^T$ and $\overline{\vec{r}}_{\text{p}2} = \setlength\arraycolsep{3pt}\begin{pmatrix}L_w & y_p & z_p\end{pmatrix}^T$, with $y_p = b/2$ and $z_p = a/2$ the local coordinates of Ports 1 and 2, respectively. Similarly, the overline in $\overline{\vec{r}}_n$ indicates that the element coordinates are local to the waveguide, i.e., $0<x \leq L_w$, $0<y\leq b$, $0 < z \leq a$. Also, $G_w(\cdot)$ is the corresponding component of the dyadic Green's function inside a rectangular waveguide, evaluated as \cite[Eq. (33)]{Williams2022}
\begin{align}
    &G_w(\vec{r}_n,\vec{r}_p) = \frac{-k_x\sin\left(\frac{\pi z_n}{a}\right)\sin\left(\frac{\pi z_p}{a}\right)}{a b k^2 \sin(k_x L_w)} \notag \\
    &\times\Big[\cos(k_x(x_p+x_n-L_w)) + \cos(k_x(L_w-|x_n-x_p|))\Big]. \label{eq:Gw}
\end{align}
This function applies for a finite waveguide terminated in a metallic wall (c.f. \cite{Li1995}), encompassing the case of zero magnetic current at every port (with the load port thus becoming a short circuit), needed to derive the mutual admittances.

\subsubsection{$\mat{Y}_\textnormal{rr}$ and $\mat{Y}_\textnormal{ll}$} these matrices represent the self-admittance at each port (coupling between a port and itself), being therefore diagonal and straightforwardly obtained as
\begin{align}
    \mat{Y}_\textnormal{rr} &= Y_\text{rr}\mat{I}_{N_w},  & Y_\text{rr} &= il_m^2\omega\varepsilon G_w\left(\overline{\vec{r}}_{\text{p}1},\overline{\vec{r}}_{\text{p}1}\right), \label{eq:Y_rr}\\
    \mat{Y}_\textnormal{ll} &= Y_\text{ll}\mat{I}_{N_w},  & Y_\text{ll} &= il_m^2\omega\varepsilon G_w\left(\overline{\vec{r}}_{\text{p}2},\overline{\vec{r}}_{\text{p}2}\right).
\end{align}

\subsubsection{$\mat{Y}_\textnormal{lr}$} similarly, $\mat{Y}_\textnormal{lr}$ captures the interaction between Port 1 and 2 at each waveguide. Then, 
\begin{align}
    \mat{Y}_\textnormal{lr} &= Y_\text{lr}\mat{I}_{N_w}, &  Y_\text{lr}&= il_m^2\omega\varepsilon G_w(\overline{\vec{r}}_{p2},\overline{\vec{r}}_{p1}).
\end{align}

\subsubsection{$\mat{Y}_\textnormal{ss}$} its main diagonal contains the self-admittances of the radiating elements, while the off-diagonal entries represent the mutual coupling. When an element is excited, it scatters the field both out of and into the waveguide, and hence coupling through the air and the waveguide needs to be considered. Specifically, the entries of $\mat{Y}_\text{ss}$ are given by
\begin{align}
    (\mat{Y}_\text{ss})_{n,n'} =& i2l_m^2\omega\varepsilon_0 G_a(\vec{r}_{n},\vec{r}_{n'}) + \notag \\
    &\begin{cases}il_m^2\omega\varepsilon G_w(\overline{\vec{r}}_n,\overline{\vec{r}}_{n'}), &  \text{\small $n, n'$ same waveguide} \\ 
    0 & \text{\small $n,n'$ different waveguide}\end{cases},
\end{align}
where $G_a(\cdot)$ is the $z$-to-$z$ component of the Green's function in free space \cite[Eq. (39)]{Williams2022}
\begin{align}
G_{a}\left(\vec{r}_n,\vec{r}_{n'}\right) =  \left( \tfrac{R^2 - \Delta z^2}{R^2} - i\tfrac{ R^2-3\Delta z^2}{R^3 k_0}-\tfrac{ R^2-3\Delta z^2}{R^4 k_0^2}\right) \frac{e^{-i k_0 R}}{4 \pi R}, \label{eq:G_a}
\end{align}
with $R = \|\vec{r}_n - \vec{r}_{n'}\|_2$, and $\Delta z = z_n - z_{n'}$. The diagonal elements of $\mat{Y}_\text{ss}$ are similarly obtained as
\begin{equation}
    (\mat{Y}_\text{ss})_{n,n} = \lim_{\vec{r}_n \rightarrow \vec{r}_{n'}}il_m^2\omega\left( 2\varepsilon_0 G_a(\vec{r}_n,\vec{r}_{n'}) + \varepsilon G_w(\overline{\vec{r}}_{n},\overline{\vec{r}}_{n})\right).
\end{equation}
As discussed in \cite{Williams2022}, the limit of $\real{G_a(\cdot)}$ diverges, and thus the imaginary part of the self-admittance is not defined. This is a consequence of the infinitesimal magnetic dipole model, as any real antenna element has a defined self-admittance. In our model, as we have an arbitrary load admittance $Y_{\text{s},n}$ attached to each element that can absorb an imaginary term, we deliberately ignore the singularity, and hence define the self-admittance as 
\begin{align}
    (\mat{Y}_\text{ss})_{n,n} \triangleq& \lim_{\vec{r}_n \rightarrow \vec{r}_{n'}}-2l_m^2\omega  \varepsilon_0\imag{G_a(\vec{r}_n,\vec{r}_{n'})} \notag \\
    &+il_m^2\omega\varepsilon G_w(\overline{\vec{r}}_{n},\overline{\vec{r}}_{n}) \notag 
    \\ =
    &\frac{l_m^2k_0\omega \epsilon_0}{3\pi} + il_m^2\omega\varepsilon G_w(\overline{\vec{r}}_{n},\overline{\vec{r}}_{n}).
\end{align}

Notice that the waveguide local coordinates $\overline{\vec{r}}_n$ are used in $G_w(\cdot)$, but not in $G_a(\cdot)$. Also, a factor of $2$ scales the latter since we consider that the \gls{DMA} is placed on a metallic plane. 

\subsubsection{$\mat{Y}_\textnormal{dd}$} \label{subsubsec:Ydd} Analogous to the previous case, the diagonal elements are the self-admittance of the devices communicating with the \gls{DMA}, while the off-diagonal elements capture the mutual coupling. We then have
\begin{align}
    (\mat{Y}_\text{dd})_{m,m'} =& il_m^2\omega \varepsilon_0 G_a(\vec{r}_m,\vec{r}_{m'}), \\
    (\mat{Y}_\text{dd})_{m,m} =& \frac{l_m^2k_0\omega \varepsilon_0}{6\pi},
\end{align}
for $m = 1,\dots,M$ where, again, we have neglected the singularity in the self-admittance. Note that, since $G_a(\cdot)$ is inversely proportional to the distance between devices $m$ and $m'$, $\mat{Y}_\text{dd}$ will be approximately diagonal unless the devices are close by (a distance of a few wavelengths suffices). 

\subsubsection{$\mat{Y}_\textnormal{ds}$} As stated above, this matrix captures the interaction between the radiating elements and the devices, hence representing the wireless propagation channel. Thus, its entries can be determined by measurements, channel simulators, or theoretical models. For exemplary purposes, a general channel model is provided in Appendix \ref{app:ChannelModel}. 

\subsubsection{System design parameters $\mat{Y}_\textnormal{s}$, $Y_\textnormal{g}$, $Y_\textnormal{l}$ and $Y_\textnormal{d}$} rather than mutual admittances, these matrices represent load admittances at the radiating elements, Ports 1, Ports 2 and devices, respectively. Therefore, their value is arbitrary and will depend on the system design and implementation constraints. Importantly, $\mat{Y}_\textnormal{s}$ characterizes the radiation properties of the elements (and their losses) and should be tunable to attain desired reconfigurability.

%------------------------------------------------------------------------------------------------------------------
% PIN BASED DMA FAS
%------------------------------------------------------------------------------------------------------------------
\section{P-i-n diode based DMA implementation for FAS}
\label{sec:FAS_DMA_Model}
The general model previously introduced for \glspl{DMA} does not impose any constraints on the number of waveguides, the element configuration, or their load admittances. Therefore, it can readily be adapted to represent an electronically reconfigurable fluid antenna implementation. To that end, consider the original baseline concept of \gls{FAS} where a one-dimensional device is fed by a single \gls{RF} chain \cite{Wong2022_BruceLee, Wong2022, Ramirez2024}, which translates into a single waveguide in the \gls{DMA} ($N_w = 1$). Note that, by stacking several waveguides, we can achieve planar implementations with multiple \gls{RF} chains, closer to the more sophisticated \gls{FAS} designs sketched in \cite{Wong2024}.

%----------------------------------------------
% ADMITTANCES EXPRESSIONS
%----------------------------------------------
\subsection{Modeling the radiating elements}

\begin{figure}[t]
    \centering
    \includegraphics[width=0.9\linewidth]{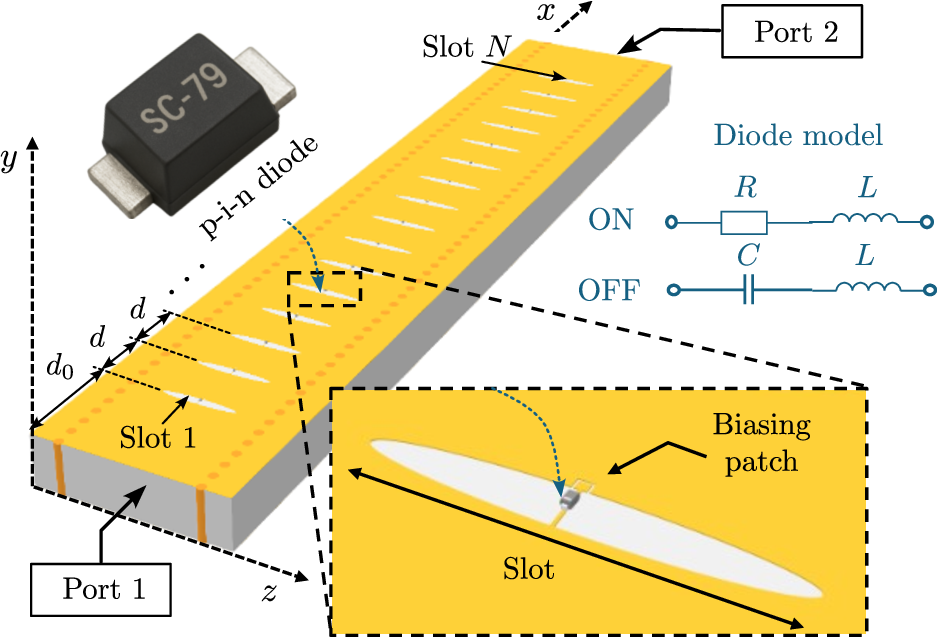}
    \caption{Baseline FAS implementation using a DMA controlled by p-i-n diodes.}
    \label{fig:DMA_FAS}
\end{figure}

Again, without loss of generality, we consider that the \gls{RF} chain (both in transmission and reception modes) is connected to Port 1 ($x = 0$), while Port 2 is an arbitrary load $Y_\text{l}$. The radiating elements are simple elliptical slots on the upper face of the waveguide ($y = b$), where the width of the slot ($x$ direction) is much smaller than the wavelength, so that the slot fundamentally behaves as a magnetic dipole. The reconfigurability of the device is achieved by attaching a \mbox{p-i-n} diode to each radiating element, as depicted in Fig. \ref{fig:DMA_FAS}. When the diode is in direct polarization (ON state), the current that flows through effectively short-circuits the slot, which thus becomes electrically invisible. In turn, in reverse polarization (diode in OFF state), the slot radiates normally. By changing the configuration of the diodes, different radiation patterns are generated, varying the way the different impinging waves are weighted and hence changing the equivalent baseband signal seen at the \gls{RF} chain. If the number of configurations is sufficiently large, a \gls{FAS} is emulated, similarly to the pixel-based antenna design in \cite{Zhang2024}. 

To capture the p-i-n diode behavior in the circuital model, we set $(\mat{Y}_\text{s})_{n,n} = Y_\text{rad}\in\mathbb{C}$ if the $n$-th diode is in reverse polarization (the $n$-th element radiates), and  $(\mat{Y}_\text{s})_{n,n} \rightarrow \infty $ if the element is short-circuited. $Y_\text{rad}$ is an arbitrary admittance characterizing the radiating properties (and the losses) of the slot when this is excited. Note that setting $(\mat{Y}_\text{s})_{n,n} \rightarrow \infty $ is equivalent to removing the $n$-th row and column of the term $(\mat{Y}_\text{s}+\widetilde{\mat{Y}}_\text{ss})^{-1}$ in \eqref{eq:jd_tx} and \eqref{eq:jr_rx}-\eqref{eq:jl_rx}, i.e., equivalent to turning off the corresponding radiating element. 

%----------------------------------------------
% FIELDS AND POWERS
%----------------------------------------------
\subsection{Power balance and radiated fields}
\label{sec:PowerBalance}
With the modeling in Section \ref{sec:DMA_Model_General} and the diode configuration presented before, we can now derive relevant metrics that are of interest when characterizing the \gls{DMA}-based \gls{FAS}. First, it is important to calculate the transmitted/received power, which is critical in communications and often overlooked in simplified models. In transmission, the overall \textit{supplied} power (by the \gls{RF} chain) is given by \cite{Pozar2012}
\begin{equation}
    P_s = \frac{1}{2}\real{\vec{j}_\text{g}^HY_\text{g}\mat{I}_{N_w}\vec{j}_\text{g}} \overset{(a)}{=} \frac{1}{2}|j_\text{g}|^2\real{Y_\text{g}},
\end{equation}
where in $(a)$ we have specialized for $N_w = 1$. Out of $P_s$, only a fraction of power is transmitted (delivered) to the waveguide, depending on the input admittance seen by the generator. This transmitted power is easily calculated as
\begin{equation}
    P_r = \frac{1}{2}\real{\vec{j}_\text{r}^H\vec{v}_\text{r}} = \frac{1}{2}|j_\text{r}|^2\real{{Y}_\text{p}}, \label{eq:Pr}
\end{equation}
where the second equality uses $N_w = 1$ and\footnote{We drop the matrix notation as $N_w = 1$ and hence $\mat{Y}_\text{p}=Y_\text{p}$.} ${v}_\text{r}={Y}_\text{p}{j}_\text{r}$. From \eqref{eq:Rel_jr_jg}, we see that $j_\text{r} = (1+\Gamma)j_\text{g}$, where $\Gamma = (Y_\text{g}-Y_\text{p})(Y_\text{g}+Y_\text{p})^{-1}$ is the reflection coefficient at the waveguide input, and therefore the relation $P_r = (1-|\Gamma|^2)P_s$ holds. Note that, as proved in \cite{Nossek2024}, modeling in terms of admittances/impedances and scattering parameters is equivalent, and thus $\Gamma$ equals the $S_{11}$ parameter.

Similarly, the power dissipated in the slots and the load is given by $P_\text{slot} = \frac{1}{2}\vec{j}_\text{s}^H\real{\mat{Y}_\text{s}}\vec{j}_\text{s}$ and $P_l=\frac{1}{2}|j_\text{l}|^2\real{{Y}_\text{l}}$ with
\begin{align}
    \vec{j}_\text{s} =& -(\mat{Y}_\text{s}+\widetilde{\mat{Y}}_\text{ss})^{-1}\widetilde{\mat{Y}}_\text{sr}{j}_\text{r}, \\
    j_\text{l} =& (\mat{Y}_\text{l}+\mat{Y}_\text{ll})^{-1}\left[\mat{Y}_\text{sl}^T(\mat{Y}_\text{s}+\widetilde{\mat{Y}}_\text{ss})^{-1}\widetilde{\mat{Y}}_\text{sr}-\mat{Y}_\text{lr}\right]j_\text{r},
\end{align}
and the \textit{radiated} power is calculated as \begin{equation}
    P_\text{rad} = \frac{1}{2}\vec{j}_\text{s}^H\real{\mat{Y}_\text{ss}}\vec{j}_\text{s}. \label{eq:Prad}
\end{equation}
As the waveguides are considered lossless, power conservation law enforces $P_r = P_\text{rad} + P_\text{slot}+ P_l$. 

In turn, when using the \gls{DMA} as a receiver, the interest lies in characterizing the power received at each port. In such case, the power received at the \gls{RF} chain is $P_\text{rx} = \frac{1}{2}|j_\text{r}|^2\real{Y_\text{g}}$ with $j_\text{r}$ as in \eqref{eq:jr_rx}, while $P_l$ and $P_\text{slot}$ are calculated as in the transmitting case.

Beyond power balance, the DMA-based \gls{FAS} implementation should ultimately be able to change the radiation pattern, emulating a physically-moving antenna. A sensible goal is then to obtain analytical expressions for the radiated field and the radiation pattern. Leveraging the infinitesimal dipole modeling of the \gls{DMA} slots, we can compute the radiated field (off the waveguide) as \cite[Eq. (29)]{Williams2022}
\begin{equation}
    \vec{h}_\text{rad}(\vec{r}) = -i2l_m\omega\epsilon_0\sum_{n=1}^N\mat{G}_{a}(\vec{r},\vec{r}_n)\hat{\vec{z}}(\vec{j}_\text{s})_n, \label{eq:RadiatedField}
\end{equation}
where $\hat{\vec{z}}$ is the unitary vector along the $z$-direction, and 
\begin{equation}
    \mat{G}_{a}(\vec{r},\vec{r}') = \frac{1}{4\pi}\left(\frac{e^{-ik_0\|\vec{r}-\vec{r}'\|_2}}{\|\vec{r}-\vec{r}'\|_2}\mat{I}_3 + \frac{1}{k_0^2}\nabla\nabla^T\frac{e^{-ik_0\|\vec{r}-\vec{r}'\|_2}}{\|\vec{r}-\vec{r}'\|_2}\right),
\end{equation}
with $\nabla$ being the vector differential operator. In far field, only the $z$-to-$z$ component of $\mat{G}_{a}(\cdot)\hat{\vec{z}}$ given in \eqref{eq:G_a} is non-zero, and therefore the field can be approximated by $\vec{h}_\text{rad}(\vec{r})\rvert_\text{ff}\approx \begin{pmatrix}0 & 0 & h_z(\vec{r})\end{pmatrix}^T$ with
 \begin{equation}
     h_z(\vec{r}) = -i2l_m\omega\varepsilon_0\sum_{n=1}^N\left(\frac{R^2-\Delta z^2}{R^2}\right)\frac{e^{-ik_0R}}{4\pi R}(\vec{j}_\text{s})_n. \label{eq:Hfarfield}
 \end{equation}
Finally, the radiation pattern (directivity) is easily calculated from the radiated field as
\begin{equation}
    D(\vec{r}) = \frac{\eta}{2}\vec{h}_\text{rad}(\vec{r})^H\vec{h}_\text{rad}(\vec{r})\frac{4\pi \|\vec{r}\|_2^2}{P_\text{rad}}, \label{eq:Directivity}
\end{equation}
with $\eta = 120\pi$ and $P_\text{rad}$ as in \eqref{eq:Prad}. Note that the last term is the inverse of the directivity achieved by an isotropic antenna radiating the same total amount of power as the \gls{DMA}.
%----------------------------------------------
% CST VALIDATION
%----------------------------------------------
\subsection{Full-wave validation}

\label{sec:fullwaveValidation}

\subsubsection{Specific design and simulation details} The proposed circuital model for \gls{DMA}-based \gls{FAS} is intended for analytical purposes, easing the theoretical characterization of realistic implementations of \gls{FAS} and serving as a tool to explore its full potential. The model accounts for all the relevant electromagnetic phenomena inherent to the radiating structure, hence being representative of practical implementations, and can be tuned to analyze specific prototypes, provided the radiating elements fit into the magnetic dipole framework. To illustrate this, we consider the design of a \gls{DMA} using a \gls{SIW} structure \cite{yan2004simulation}, as illustrated in Fig. \ref{fig:DMA_FAS}. This technology provides a simple and cost-efficient solution, benefiting from standard \gls{PCB} manufacturing processes for mass production. Although the \gls{SIW} structure consists of lateral metallic vias, it preserves the propagation behavior of a conventional rectangular waveguide, and thus the proposed model is completely applicable.

\begin{table}[t]
    \renewcommand{\arraystretch}{1.2}
     \centering
       \caption{Simulation parameters. Frequency is in GHz, and physical measurements are in millimeters.}
       \label{tab:FullWaveParam}
  \begin{tabular}{c|c||c|c}
        \textbf{Parameter}  & \textbf{Value} & \textbf{Parameter} & \textbf{Value}\\ \hline\hline
        Frequency & $f = 2.4$ & Rel. Permittivity & $\varepsilon_r = 3.55$ \\ \hline 
        Wvg. height ($z$) & $a = 58$ & Wvg. width ($y$) & $b = 23.2$ \\ \hline
        Wvg. length ($x$) & $L_w = 475$ & $\#$ slots & $N=16$ \\ \hline
        Slot major axis & 33 & Slot minor axis & 4 \\ \hline
        Location Slot 1 & $x = 50$ & Separation slots & $d = 25$ \\ \hline
        \multicolumn{2}{c||}{Load admittance} & \multicolumn{2}{c}{$Y_\text{rad} = 3.7\cdot 10^{-5}-i0.0037$} \\
        \hline\hline
        \textbf{Configurations} & \multicolumn{3}{|c}{\textbf{Active slots}} \\ \hline \hline
        Conf. 1 & \multicolumn{3}{|c}{$\begin{matrix}2 & 3 & 5 & 6 & 10 & 11 & 14 & 15\end{matrix}$} \\ \hline
        Conf. 2 & \multicolumn{3}{|c}{$\begin{matrix}1 & 2 & 3 & 7 & 8 & 13 & 14 & 15\end{matrix}$} \\ \hline
        Conf. 3 & \multicolumn{3}{|c}{$\begin{matrix}1 & 4 & 5 & 8 & 11 & 12 & 14 & 15\end{matrix}$}
  \end{tabular}
\end{table}

Following the notation in Fig. \ref{fig:DMA_schematic}, the waveguide has a width of $a=58$ mm along the $z$-axis, a height of $b = 23.2$ mm along the $y$-axis, and a length of $L_w = 475$ mm along the $x$-axis. The substrate has a relative permittivity of $\varepsilon_r = 3.55$, and the frequency of operation is $f = 2.4$ GHz. $N=16$ elliptical slots are located on the top layer ($y=b$), with major and minor axes of 33 mm and 4 mm, respectively. The slot centers are placed along the $x$-axis direction at $z = a/2$, with the first slot located at a distance $d_0 = 50$ mm from Port 1 ($x=0$), and the other slots equally separated from each other with distance $d = 25$ mm. The commercial p-i-n diode SMP1345-079LF is considered along with corresponding values of the circuit model for the ON and OFF states~\cite{Pin_diode_model}. Since the diode is smaller than the slot minor axis, metallic edges are introduced around each aperture to center the diode physically over the slot. Besides, to ensure proper biasing of the diode, a small patch is isolated on one side of the edges (see Fig. \ref{fig:DMA_FAS}). This design keeps the radiation properties of the slots while rendering a realistic structure with a localized isolated region and a continuous ground plane. As explained before, the diode remains in the OFF state when no bias current is applied, enabling the slot radiation at its resonance frequency. Under forward bias, though, the diode switches to the ON state, presenting a low impedance that effectively suppresses the slot radiation (short-circuit). Full-wave simulations of the structure, assuming lossless dielectric, are carried out using CST Microwave Studio, and a summary of the simulation parameters is provided in Table \ref{tab:FullWaveParam}. 

\subsubsection{Fitting the theoretical model} Regarding the application of the model, note that all the mutual admittances in Section \ref{sec:DMA_Model_General} are directly computed by introducing the corresponding structure parameters into the analytical expressions, where the coordinates of the slot centers are directly used as $\vec{r}_n$ (position of magnetic dipoles). Hence, it only remains to specify the dipole length $l_m$ and the load admittances $\mat{Y}_\text{s}$. For the former, we choose $l_m$ such that the waveguide self-admittance equals the simulated one in CST $Y_0^\text{cst}$, leading to\footnote{The theoretical self-admittance can be obtained as the input admittance $Y_\text{rr}$ at Port 1 when using a semi-infinite waveguide of the same dimensions. Hence, it is calculated as \eqref{eq:Y_rr} but replacing $G_w(\cdot)$ by the corresponding component of \cite[Eq. (9)]{Li1995}.}
\begin{equation}
    l_m = \sqrt{Y_0^\text{cst}\frac{abk^2}{2\omega\varepsilon k_x}}. \label{eq:lm}
\end{equation}

On the other hand, the matrix $\mat{Y}_\text{s}$ characterizes the radiation properties of the slots at the target frequency. As stated before, we have $(\mat{Y}_\text{s})_{n,n}=Y_\text{rad}$ for the \gls{DMA} in Fig. \ref{fig:DMA_FAS} if slot $n$ is active, and $(\mat{Y}_\text{s})_{n,n}\rightarrow \infty$ if the $n$-th slot is off (diode in forward bias). Thus, the objective is to characterize $Y_\text{rad}$, which can be obtained by trial and error (based on full-wave simulations or empirical measurements) or by fitting to some magnitude. We here rely on the latter approach, proposing the scattered field within the waveguides as baseline. Specifically, we solve
\begin{equation}
    \min_{Y_{\text{rad}}} \quad \frac{1}{Q}\sum_{q=1}^Q |h_{z,w}(\vec{r}_q) - h^\text{cst}_{z,w}(\vec{r}_q)|, \label{eq:OptYs}
\end{equation}
where $\{\vec{r}_q\}_{q=1}^Q$ is a set of points along the waveguide center ($z=a/2$, $y=b/2$), $h^\text{cst}_{z,w}(\vec{r}_q)$ is the transversal component of the magnetic field ($z$ direction) obtained by full-wave simulations, and  $h_{z,w}(\vec{r}_q)$ is the theoretical component of the field, computed analogously to \eqref{eq:RadiatedField} as
\begin{align}
    h_{z,w}(\vec{r}_q) =& -il_m\omega\varepsilon\Big(\underbrace{G_w(\vec{r}_q,\bar{\vec{r}}_{\text{p}1})j_\text{r}}_{\text{Port 1 contribution}} + \underbrace{G_w(\vec{r}_q,\bar{\vec{r}}_{\text{p}2})j_\text{l}}_{\text{Port 2 contribution}} \notag \\
    &+ \underbrace{\sum_{n=1}^N G_w(\vec{r}_q,\bar{\vec{r}}_{n})(\vec{j}_\text{s})_n}_{\text{Slots contribution}}\Big). \label{eq:HfieldWaveguide}
\end{align}

\begin{remark}
    The model calibration and the \gls{DMA} analysis in general can be done for any value of $l_m$, which acts as a simple scaling constant. In fact, if $\mat{Y}_\textnormal{s}$ is fitted for $l_m = 1$, then it is valid for any other $l_m$ by simply scaling $\mat{Y}_\textnormal{s}\leftarrow l_m^2\mat{Y}_\textnormal{s}$. This is easily proved by applying this scaling in, e.g.,  the radiated field in \eqref{eq:RadiatedField} and the waveguide field \eqref{eq:HfieldWaveguide}, where $l_m$ cancels out. Hence, although we use the value in \eqref{eq:lm}, the validation results are exactly the same for any other $l_m$. 
\end{remark}

\subsubsection{Validation results} We consider three specific configurations for the designed \gls{DMA}-based \gls{FAS}, as specified in Table \ref{tab:FullWaveParam}. Model calibration is done for Configuration 3 by numerically solving \eqref{eq:OptYs} for an arbitrary set of points $\{\vec{r}_q\}$ when the waveguide is terminated in a matched load $Y_\text{l}=Y_0^\text{cst}$, and $l_m$ is chosen as in \eqref{eq:lm}. Importantly, once calibrated, the values of $l_m$ and $Y_\text{rad}$ remain constant for any other configuration, and should be recalculated only after modifying any physical parameter in the design. The other two configurations are used next to evaluate the modeling.

\begin{figure}[t]
    \centering
    \includegraphics[clip, trim={0 2cm 0 2.5cm},width=1\linewidth]{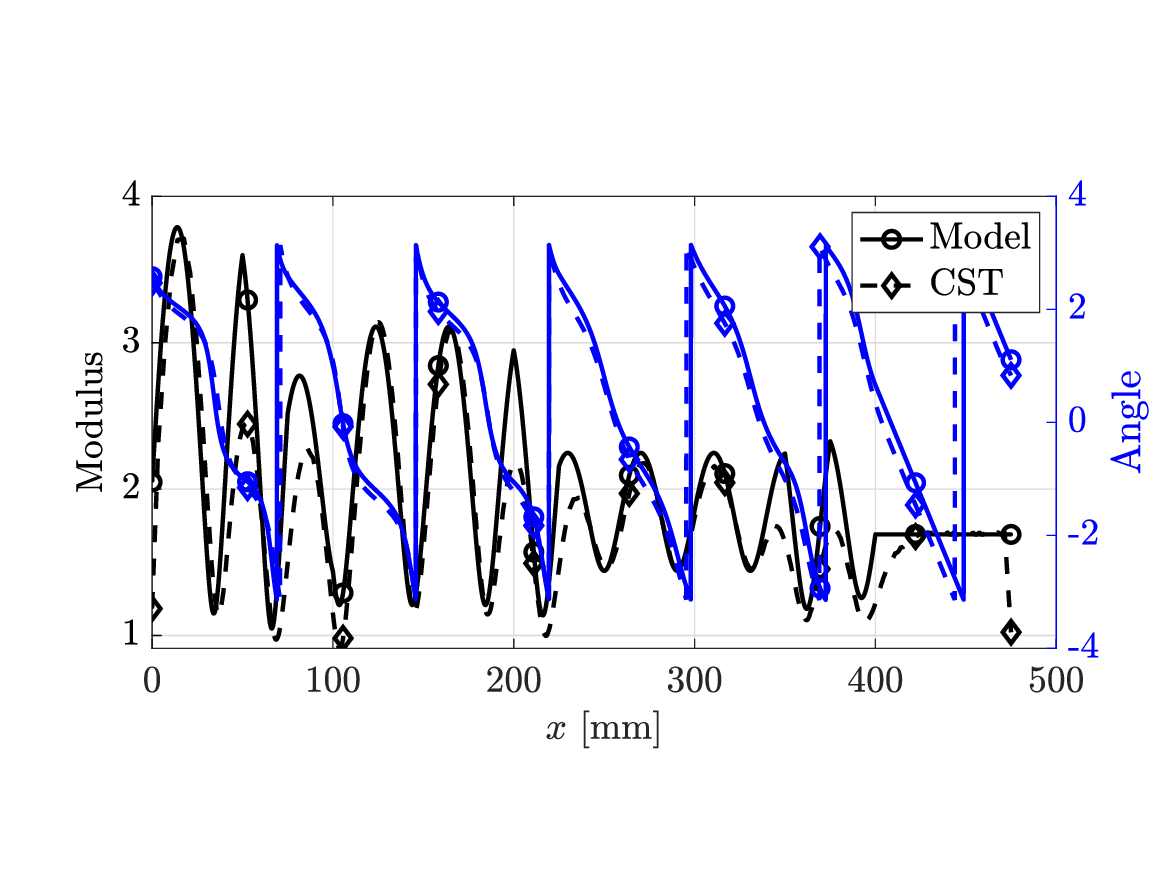}
    \caption{Transversal ($z$) component of magnetic field inside waveguide (on the axis along $y = b/2$ and $z = a/2$) for DMA configuration 1 when ended up in a matched load. Solid lines correspond to the theoretical model, while dashed lines correspond to CST full-wave simulations. }
    \label{fig:Field}
\end{figure}

The first result is shown in Fig. \ref{fig:Field}, which plots the transversal component of the magnetic field along the center of the waveguide for configuration 1 and $Y_\text{l} = Y_0^\text{cst}$. We observe a relatively good agreement, both in terms of magnitude and phase, between the simulated and the theoretical field, being the latter calculated as in \eqref{eq:HfieldWaveguide}. Although the field is not strictly necessary to characterize \gls{FAS}, it is indeed interesting to see the ability of the model to capture the propagation inside the waveguides. 

\begin{figure}[t]
    \centering
    \includegraphics[width=0.9\linewidth]{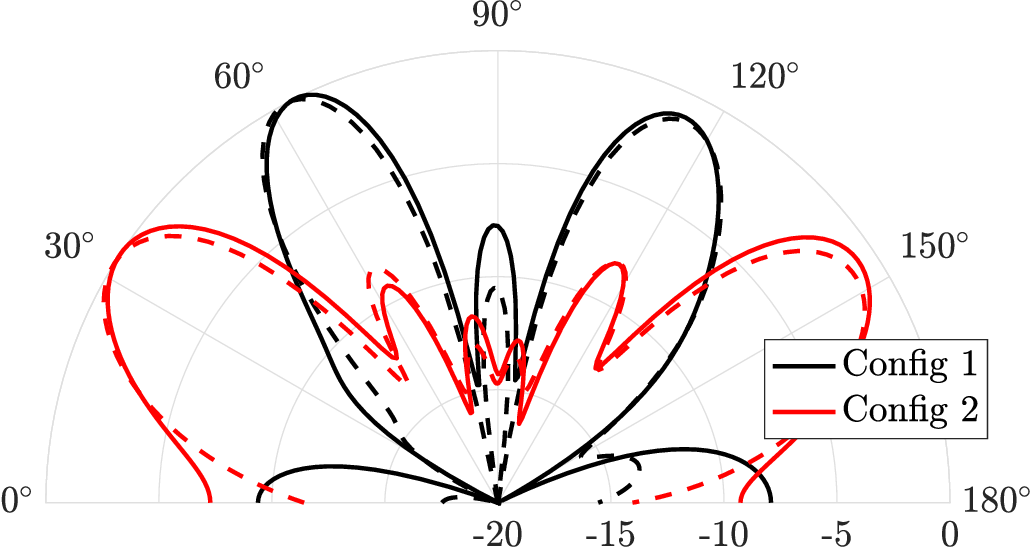}
    \caption{Radiation pattern cut in azimuth plane for DMA configurations 1 and 2 when ended up in a short-circuit. Solid lines correspond to the theoretical model, while dashed lines correspond to CST full-wave simulations. The patterns are normalized and depicted in dB.}
    \label{fig:Rad_Pattern_short}
\end{figure}

Moving on towards \glspl{FAS}-oriented metrics, Figs. \ref{fig:Rad_Pattern_short} and \ref{fig:Rad_Pattern_load} show the azimuth cut of the radiation pattern (see Fig. \ref{fig:coordinates} in Appendix \ref{app:ChannelModel} for angle definition) generated by the proposed design when the waveguide is terminated in a short-circuit (Fig. \ref{fig:Rad_Pattern_short}) and a matched load (Fig. \ref{fig:Rad_Pattern_load}). The theoretical radiation patterns are calculated by using \eqref{eq:Directivity} for points $\vec{r}$ in the $xy$-plane ($\theta = \pi/2$, $\phi \in[0, \pi)$), and then normalized to 0 dB, with the same normalization applied to the simulated radiation patterns. We see a good match between theoretical and simulated results in both figures, where the model is shown to capture the main lobes as well as the most side lobes fairly accurately. The mismatch is larger as we approach endfire directions ($\phi = 0^\circ$ or $\phi = 180^\circ$), i.e., closer to the plane containing the \gls{DMA}. This is due to the infinite \gls{PEC} plane assumed in the model, which is naturally not present in the simulation, and other edge effects. Remark that each configuration (with distinct radiation pattern) can be seen as a \textit{port} in the usual \gls{FAS} nomenclature. 

\begin{figure}[t]
    \centering
    \includegraphics[width=0.9\linewidth]{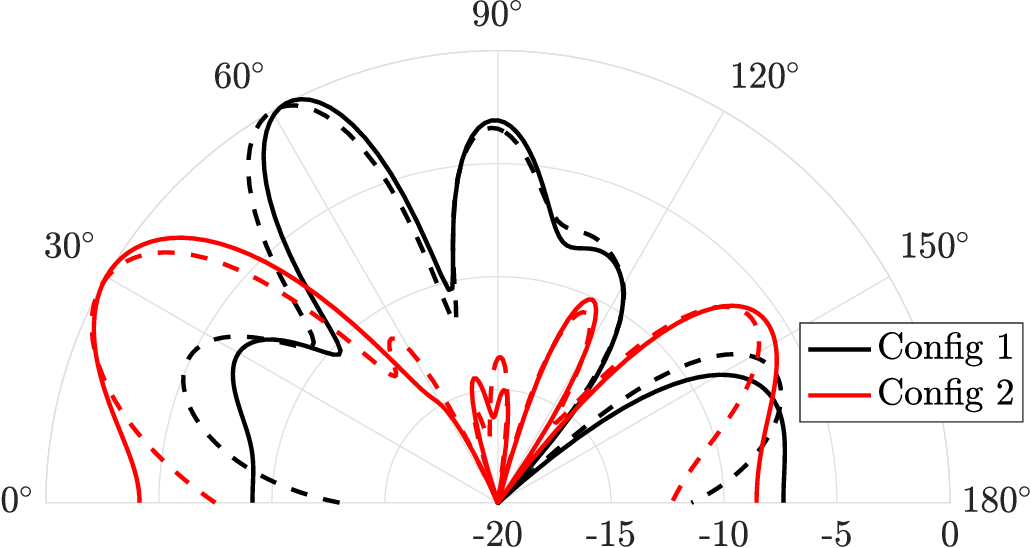}
    \caption{Radiation pattern cut in azimuth plane for DMA configurations 1 and 2 when ended up in a matched load. Solid lines correspond to the theoretical model, while dashed lines correspond to CST full-wave simulations. The patterns are normalized and depicted in dB.}
    \label{fig:Rad_Pattern_load}
\end{figure}

\begin{figure}[t]
    \centering
    \includegraphics[width=1\linewidth]{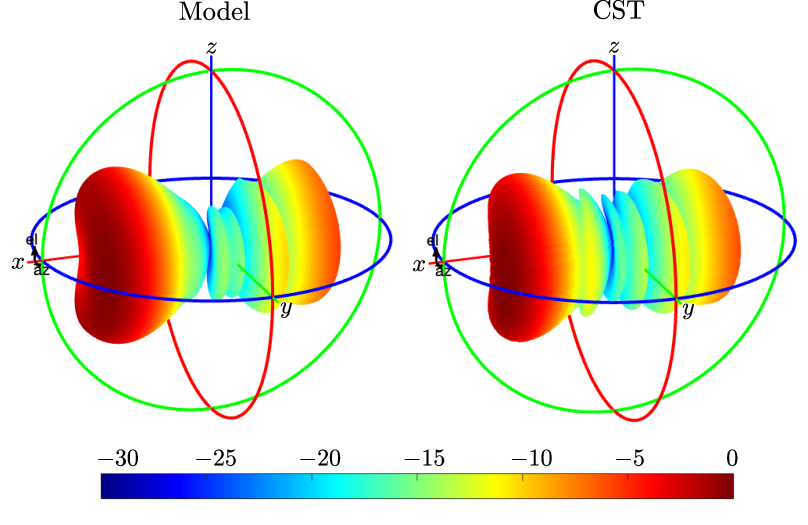}
    \caption{3D radiation pattern for configuration 2 when ended up in a matched load. The patterns are normalized and depicted in dB.}
    \label{fig:Rad_Pattern_3D}
\end{figure}

Figure \ref{fig:Rad_Pattern_3D} illustrates the 3D theoretical and simulated radiation pattern achieved by configuration 2 when the waveguide is loaded with $Y_\text{l} = Y_0$. Again, the patterns are normalized, and the theoretical result is calculated using \eqref{eq:Directivity}, but now $\vec{r}$ moves along the half-sphere containing the \gls{DMA} ( $\theta,\phi\in[0,\pi)$). Again, we observe how the model successfully captures the electromagnetic properties of the structure. The proposed modeling framework thereby postulates as a solid (yet tractable) tool for the analysis of realistic \gls{FAS} implementations.

As a final remark, it should be noted that we have restricted the design by considering single-fed point structures only, i.e., Port 2 is a simple load. However, a dual-port structure could, in principle, provide larger flexibility, allowing, e.g., for the radiation of mirrored beams across the broadside axis.

%------------------------------------------------------------------------------------------------------------------
% COMMUNICATIONS MODEL FOR FAS
%------------------------------------------------------------------------------------------------------------------
\section{Communications Model for FAS}
\label{sec:FAS_Model}
%----------------------------------------------
% RECEIVED SIGNAL
%----------------------------------------------
\subsection{Received signal model}

\label{sec:FAMA_signal_model}

Consider now the same \gls{FAMA} setup as in \cite{Wong2022, Ramirez2024}, where a \gls{BS} with $M$ independent fixed antennas simultaneously serves $M$ users, each equipped with a fluid antenna. Relating this setup with the previous circuital model, we have that each \gls{BS} antenna is represented by a single antenna device supplying a magnetic current $(\vec{j}_\text{g})_m$ with $m = 1,\dots,M$, while each user, equipped with a p-i-n diode based \gls{DMA} fluid antenna, is characterized by \eqref{eq:jr_rx}-\eqref{eq:Rel_jd_jg}. Assuming, without loss of generality, that all the fluid antennas have the same specifications, the current induced at the $m$-th user is obtained from \eqref{eq:jr_rx} and \eqref{eq:Rel_jd_jg} as
\begin{equation}
    j_{\text{r},u}^{(m)} = \frac{2Y_\text{d}}{Y_\text{g}+Y_\text{p}}\widetilde{\mat{Y}}_\text{sr}^T(\mat{Y}_\text{s}^{(u)}+\widetilde{\mat{Y}}_\text{ss})^{-1}\mat{Y}_\text{ds}^{(m)T}(Y_\text{d}\mat{I}_M + \mat{Y}_\text{dd})^{-1}\vec{j}_\text{g}, \label{eq:FAMA_jr_general}
\end{equation}
where $\mat{Y}_\text{s}^{(u)}$ indicates that user $m$ is using the \gls{DMA} configuration\footnote{$Y_\text{p}$ is also a function of the configuration $\mat{Y}_\text{s}^{(u)}$ as stated in \eqref{eq:Yp}. However, we drop this dependence from the notation for the sake of clarity.} $u$ and $\mat{Y}_\text{ds}^{(m)}$ is the wireless channel from the \gls{BS} to user $m$. Note that we have $N_w = 1$ and therefore we drop the matrix notation in $\mat{Y}_\text{p}$. 

In \gls{FAMA}, each BS antenna transmits the signal intended for one user, while signals arriving from the other \gls{BS} antennas are seen as interference. To make this distinction in \eqref{eq:FAMA_jr_general}, write $\mat{Y}_\text{ds}^{(m)} = \begin{pmatrix}\vec{y}_{\text{ds},1}^{(m)} & \vec{y}_{\text{ds},\widetilde{m}}^{(m)}& \dots & \vec{y}_{\text{ds},M}^{(m)}\end{pmatrix}^T$, where $\vec{y}_{\text{ds},\widetilde{m}}^{(m)}\in\mathbb{C}^{N\times1}$ is the channel from \gls{BS} antenna $\widetilde{m}$ to user $m$. Also, consider that the \gls{BS} antennas are separated enough so there is no mutual coupling between them, implying that $\mat{Y}_\text{dd}$ is diagonal (see Section \ref{subsubsec:Ydd}). Under admittance matching conditions---i.e., $\mat{Y}_\text{dd} = Y_\text{d}\mat{I}_M$---the induced current \eqref{eq:FAMA_jr_general} is rewritten as
\begin{align}
    j_{\text{r},u}^{(m)} =& \underbrace{\frac{1}{Y_\text{g}+Y_\text{p}}\widetilde{\mat{Y}}_\text{sr}^T(\mat{Y}_\text{s}^{(u)}+\widetilde{\mat{Y}}_\text{ss})^{-1}\vec{y}_{\text{ds},m}^{(m)}j_\text{g}^{(m)}}_{\text{desired signal}}  \notag \\
    &+\underbrace{\frac{1}{Y_\text{g}+Y_\text{p}}\widetilde{\mat{Y}}_\text{sr}^T(\mat{Y}_\text{s}^{(u)}+\widetilde{\mat{Y}}_\text{ss})^{-1}\sum_{\widetilde{m}\neq m}\vec{y}_{\text{ds},\widetilde{m}}^{(m)}j_\text{g}^{(\widetilde{m})}}_{\text{interference}}, 
\end{align}
where $j_\text{g}^{(m)} = (\vec{j}_\text{g})_m$. Finally, note that $j_\text{g}^{(m)}$ is the magnetic current supplied by \gls{BS} antenna $m$, i.e., a complex voltage supplied by some generator. We can readily identify it with the complex transmitted symbols, adopting common notation in wireless communications. Thus, denote $j_\text{g}^{(m)} = x_m\in\mathbb{C}$. Similarly, $j_\text{r}^{(m)}$ is the received complex voltage---equivalently, the received symbol---and then $j_\text{r}^{(m)} = s_m\in\mathbb{C}$. Using these relations, and defining\footnote{Please, note that as $N_w = 1$ then $\widetilde{\mat{Y}}_\text{sr}$ has dimensions $N\times 1$.} \begin{equation}
    \vec{g}_u^T= \frac{1}{Y_\text{g}+Y_\text{p}}\widetilde{\mat{Y}}_\text{sr}^T(\mat{Y}_\text{s}^{(u)}+\widetilde{\mat{Y}}_\text{ss})^{-1}, \label{eq:DMAresponse_gu}
\end{equation}
the received complex symbols are expressed as
\begin{equation}
    s_{m,u} = \vec{g}_u^T\vec{y}_{\text{ds},m}^{(m)}x_m + \sum_{\widetilde{m}\neq m}\vec{g}_u^T\vec{y}_{\text{ds},\widetilde{m}}^{(m)}x_{\widetilde{m}} + w_m, \label{eq:s}
\end{equation}
where we have added the noise term $w_m$. The above expression is in the same form of the standard signal model for \gls{FAMA} (c.f. \cite[Eq. (23)]{Ramirez2024}), representing the received complex symbol at user $m$ when the (\gls{DMA}) configuration $u$ is selected, i.e., when the fluid antenna selects `port' $u$.
\begin{remark}
\label{remark:Channel}
The equivalent channel ${h}_{u,\widetilde{m}}^{(m)} = \vec{g}_\text{u}^T\vec{y}_{\textnormal{ds},\widetilde{m}}^{(m)}$ is the inner product of two separate responses: \textit{i)} the wireless channel $\vec{y}_{\textnormal{ds},\widetilde{m}}^{(m)}$, which depends on the propagation scenario, and \textit{ii)} the \gls{DMA} response $\vec{g}_u$, which depends on the selected configuration. This highlights the difference between moving the antenna in space and moving the antenna in the beamspace. While the original concept of \gls{FAS} would modify $\vec{y}_{\textnormal{ds},\widetilde{m}}^{(m)}$ by physically moving the device, a reconfigurable antenna changes $\vec{g}_u$ while keeping the wireless channel fixed. 
\end{remark}

%----------------------------------------------
% CORRELATION
%----------------------------------------------
\subsection{Covariance of the received signal}

\label{sec:Covariance}

In \glspl{FAS}, the correlation between sampled channels at different ports is of paramount importance, fundamentally limiting the achievable performance \cite{Ramirez2024}. As pointed out in Remark \ref{remark:Channel}, under the original concept of \gls{FAS}, spatial correlation is imposed by the propagation environment and, in general, it is a function of the angle-of-arrival distribution and the separation between antenna ports. This is different in a reconfigurable antenna implementation, where instead of physical \emph{ports}, we have different metasurface (e.g., DMA) configurations playing the role of \emph{virtual ports}. These configurations change the way the impinging waves are weighted and, ultimately, the correlation between the `sampled' channels (at different configurations). An equally important remark is that the received power depends on the antenna configuration, as different choices of $\mat{Y}_\text{s}$ not only impact the equivalent channel but also the reflection coefficient. 

From \eqref{eq:s}, the covariance of the equivalent channels from \gls{BS} antenna $\widetilde{m}$ to user $m$ for configurations $u$ and $\widetilde{u}$ is given by
\begin{align}
    \text{Cov}\left[h_{u,\widetilde{m}}^{(m)}, {h}_{\widetilde{u},\widetilde{m}}^{(m)}\right] =& \mathbb{E}\left[h_{u,\widetilde{m}}^{(m)}{h}_{\widetilde{u},\widetilde{m}}^{(m)H}\right]-\mathbb{E}\left[{h}_{{u},\widetilde{m}}^{(m)}\right]\mathbb{E}\left[{h}_{\widetilde{u},\widetilde{m}}^{(m)H}\right] \notag \\
    =& \vec{g}_u^T \bm{\Sigma}_{\vec{y},\widetilde{m}}^{(m)}\vec{g}_{\widetilde{u}}^*, \label{eq:Covariance}
\end{align}
where $\bm{\Sigma}_{\vec{y},\widetilde{m}}^{(m)} = \text{Cov}[\vec{y}_{\textnormal{ds},\widetilde{m}}^{(m)},\vec{y}_{\textnormal{ds},\widetilde{m}}^{(m)}]$ is the covariance matrix of the wireless channel from \gls{BS} antenna $\widetilde{m}$ to all the radiating elements in the \gls{DMA} of user $m$, and $\vec{g}_u$ is the DMA response for configuration $u$ \eqref{eq:DMAresponse_gu}. 
Assuming each user has a codebook of $\mathcal{U}$ different configurations, the resulting covariance matrix between the $\mathcal{U}$ equivalent \gls{FAS} ports is $\bm{\Sigma}_{\widetilde{m}}^{(m)}\in\mathbb{C}^{\mathcal{U}\times \mathcal{U}}$ with entries
\begin{equation}
\left(\bm{\Sigma}_{\widetilde{m}}^{(m)}\right)_{u,\widetilde{u}} = \vec{g}_u^T \bm{\Sigma}_{\vec{y},\widetilde{m}}^{(m)}\vec{g}_{\widetilde{u}}^*. \label{eq:Covariance_FAS}
\end{equation}

\begin{remark}
    \label{remark:Correlation}
    Eq. \eqref{eq:Covariance_FAS} illustrates that the resulting covariance matrix (i.e., the effective channel spatial correlation) can be tuned, up to some degree, by the set of designed \gls{DMA} configurations. Note that the wireless channel covariance $\bm{\Sigma}_{\vec{y},\widetilde{m}}^{(m)}$ is imposed by the propagation environment, and can only be modified by varying the spacing between the radiating elements (i.e., physically altering the antenna aperture). However, while this is typically fixed for a fixed aperture or metasurface, the DMA response $\vec{g}_u$ offers the necessary degree of flexibility to modify the effective channel correlation. This raises a fundamentally important question, so far overlooked in FAS: what is the most beneficial channel correlation structure? and brings the opportunity to suitably design the spatial correlation, opening up new research avenues in the design, analysis and optimization of metasurface-based FAS.
    
    %\textcolor{magenta}{This reveals a problem that has so far been overlooked in %This opens up a problem, thus far ignored in 
    %\gls{FAS}}: the design of correlation (or covariance) matrices and analysis of the correlation structure\textcolor{magenta}{s that most benefit the system}. \textcolor{magenta}{Notably, this opens the door to new optimization strategies in the design of wireless communication systems.}
%   This reveals a problem that has so far been overlooked in FAS: the design of correlation (or covariance) matrices and the analysis of correlation structures that most benefit the system."
\end{remark}

As an illustrative example, consider a \gls{DMA}-based \gls{FAS} as in Section \ref{sec:fullwaveValidation} but with $N=20$ radiating elements. The first slot is located now at $x = 60$ mm, while the separation between slots is again $d = 25$ mm and the waveguide is terminated in a short-circuit ($Y_\text{l}\rightarrow \infty$). The overall waveguide length is thus $L_w = 595$ mm, while the rest of parameters are the same as in Table \ref{tab:FullWaveParam}. To generate a codebook, we adopt the criterion of maximizing the radiated (received) power towards (from) different angular directions, so that each codeword renders a configuration pointing towards (from) a different spatial direction. Specifically, for a set of azimuth angles $\phi\in[\phi_\text{min},\phi_\text{max})$ and $\theta$ fixed to $\pi/2$, we solve
\begin{equation}
    \max_\mat{S} |h_z(\vec{r})|, \label{eq:CodebookProblem}
\end{equation}
where $h_z(\vec{r})$ is given by \eqref{eq:Hfarfield}, and $\mat{S}$ is a selection matrix indicating the active slots in the \gls{DMA}. Since $N=20$, we consider configurations with $N_\text{active} = 10$ active elements. The above optimization problem is solved by brute force, checking all the possible configurations and selecting the optimal one. Note that, as the field and currents are all expressed in closed-form, solving \eqref{eq:CodebookProblem} only takes a few seconds. The choice of maximizing the field strength instead of the directivity in \eqref{eq:Directivity} is motivated by the insertion losses and radiation efficiency. As changing the configuration modifies the reflection coefficient ($S_{11}$ parameter), maximizing the radiated power implicitly accounts for these factors. 

\begin{figure}[t]
    \centering
    \includegraphics[width=1\linewidth]{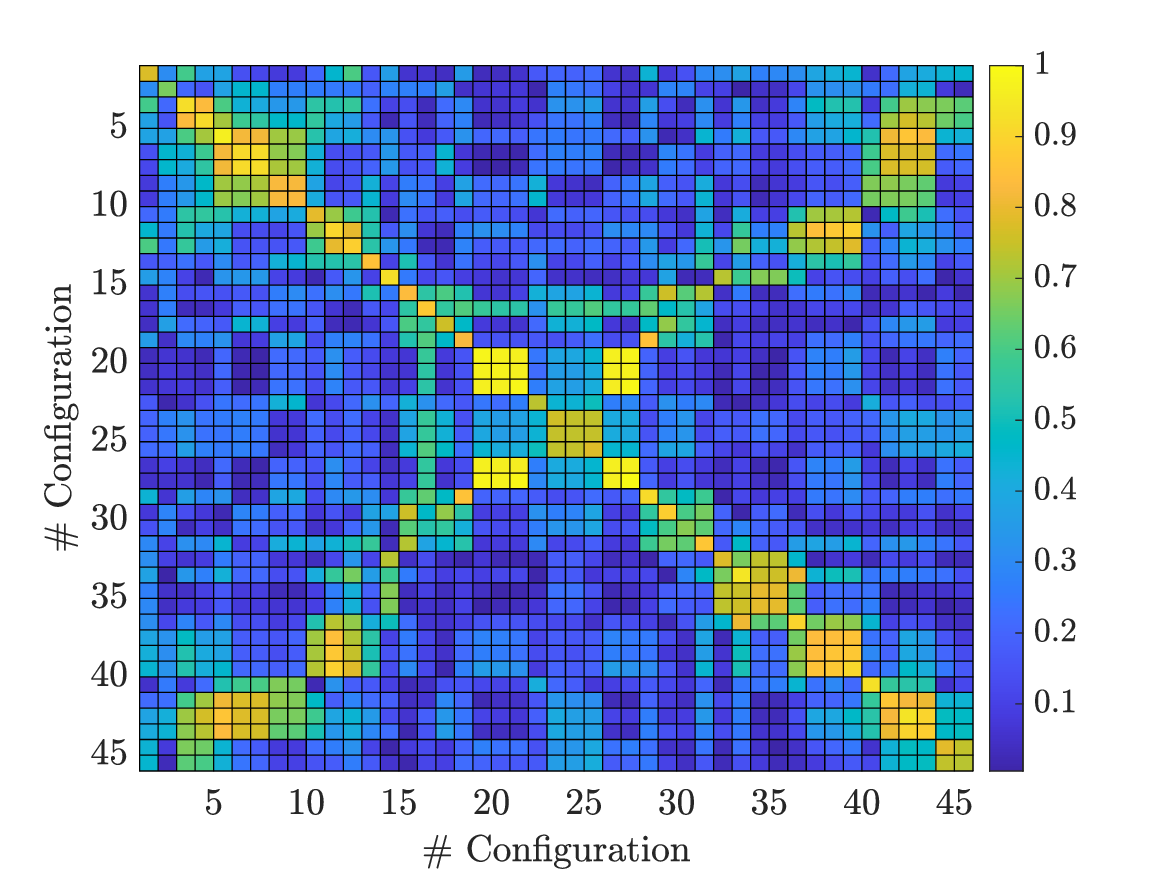}
    \caption{Modulus of the normalized covariance matrix $\bm{\Sigma}_{\widetilde{m}}^{(m)}$ in \eqref{eq:Covariance_FAS} for a codebook with 46 codewords and 3D isotropic propagation covariance in \eqref{eq:SigmaYiso}.}
    \label{fig:Covariance}
\end{figure}

Sweeping between $\phi_\text{min} = 0$ and $\phi_\text{max} = \pi$, we obtain $\mathcal{U}=46$ different configurations, where most of the configurations achieve an $S_{11}$ parameter lower than $-10$ dB, and the average radiation efficiency, calculated as $P_\text{rad}/P_s$ (see Section \ref{sec:PowerBalance}), is around $70\%$. With this codebook, the equivalent covariance matrix in \eqref{eq:Covariance_FAS} is calculated by assuming $\bm{\Sigma}_{\vec{y},\widetilde{m}}^{(m)}$ is given by \eqref{eq:SigmaYiso} (isotropic propagation), and its element-wise modulus (normalized to unity) is depicted in Fig. \ref{fig:Covariance}. As expected, we see that close-by configurations---i.e., radiation patterns pointing towards similar angular directions---experience a high correlation. Also, we can make some non-straightforward observations. First, we see that the diagonal elements are not 1, which implies that the received power varies with the configuration, as explained before. Second, the covariance structure presents a remarkable symmetry, related to the symmetry in the radiation patterns that can be observed in Fig. \ref{fig:Rad_Pattern_short}. This implies that, when terminated in a short-circuit, the \gls{DMA} symmetry translates into the beamspace, and hence one cannot point towards a direction $\phi$ without also pointing out to the mirrored direction $\pi - \phi$. Naturally, we can avoid this behavior with alternative implementations. 

\begin{figure}
    \centering
    \includegraphics[width=1\linewidth]{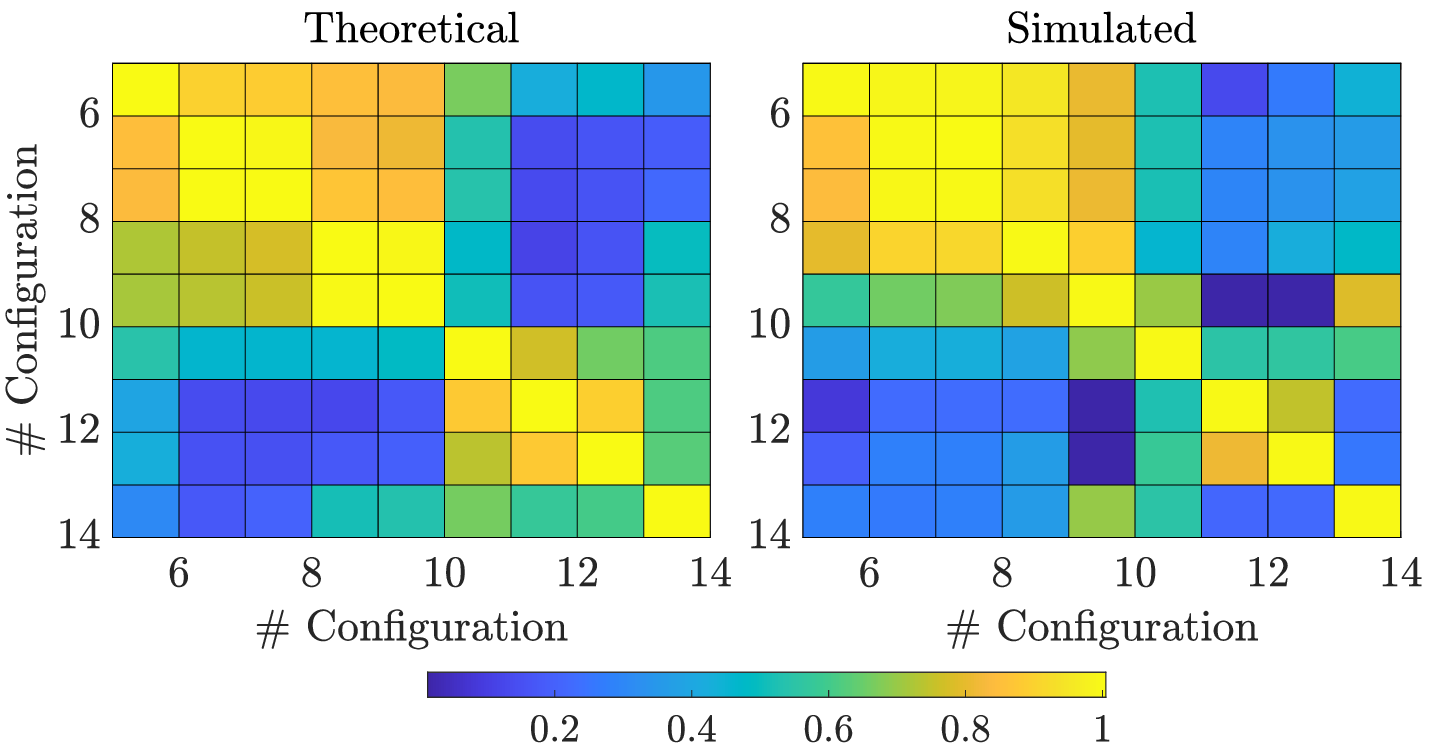}
    \caption{Modulus of the theoretical and simulated correlation coefficient for configurations 5 to 14 in Fig. \ref{fig:Covariance}.}
    \label{fig:correlation}
\end{figure}

Focusing on the ability of the theoretical model to capture similarities between configurations (radiation patterns), regardless of the attained reflection coefficient or efficiency, in Fig. \ref{fig:correlation} we compare the modulus of the theoretical correlation coefficient for a subset of exemplary configurations with the one simulated in CST Microwave Studio. Specifically, the theoretical correlation is obtained as
\begin{equation}
    \rho_{u,\widetilde{u}} = \frac{\left(\bm{\Sigma}_{\widetilde{m}}^{(m)}\right)_{u,\widetilde{u}}}{\sqrt{\left(\bm{\Sigma}_{\widetilde{m}}^{(m)}\right)_{u,{u}}}\sqrt{\left(\bm{\Sigma}_{\widetilde{m}}^{(m)}\right)_{\widetilde{u},\widetilde{u}}}},
\end{equation}
while the simulated one is calculated by (numerically) averaging the complex radiation pattern predicted by CST over the angle distribution in \eqref{eq:PDF_theta_phi}. Remarkably, we see that the model replicates the simulated correlation structure with relatively high accuracy. 

\begin{figure}[t]
    \centering    \includegraphics[trim={0 2cm 0 2.5cm},clip,width=1\linewidth]{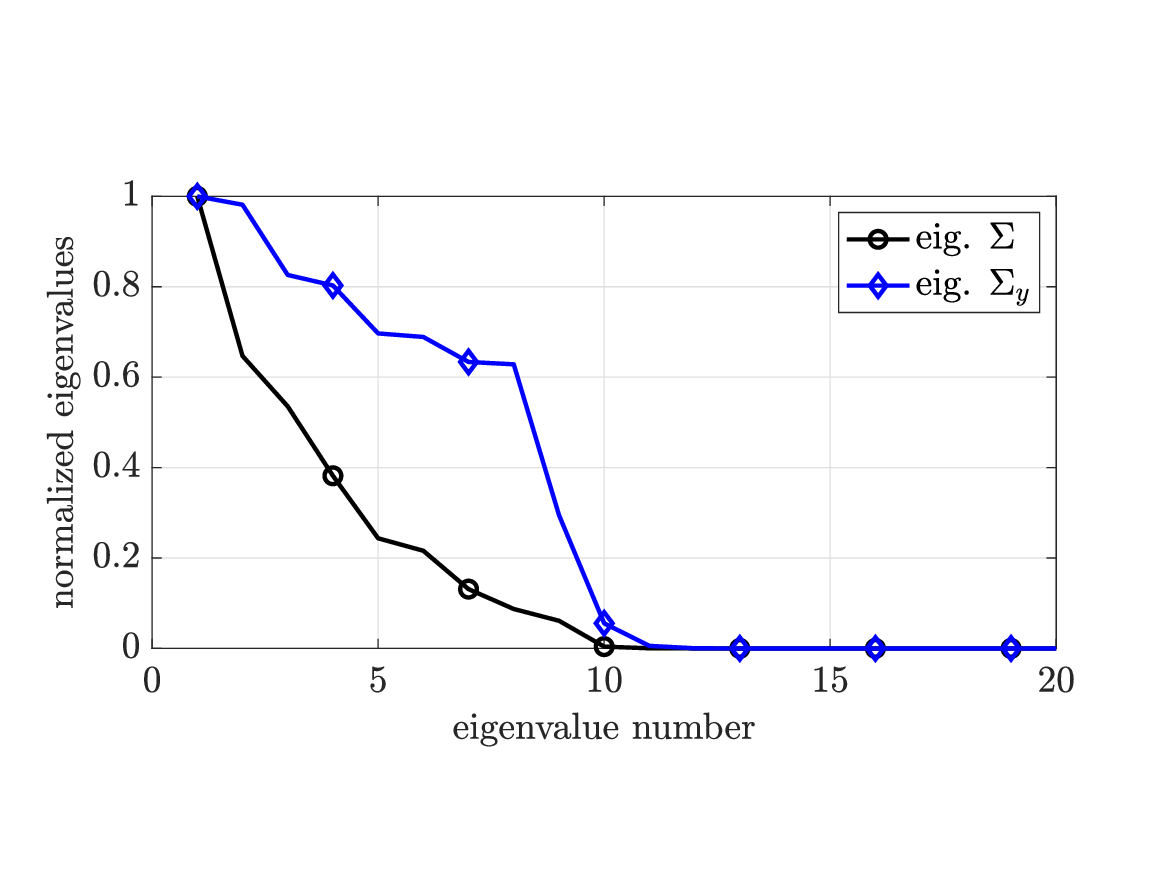}
    \caption{Normalized eigenvalues of the FAS covariance matrix $\bm{\Sigma}_{\widetilde{m}}^{(m)}$ in \eqref{eq:Covariance_FAS} vs. those of the 3D isotropic covariance matrix in \eqref{eq:SigmaYiso}.}
    \label{fig:eigCovariance}
\end{figure}

An important aspect for communications performance, as thoroughly discussed in \cite{Ramirez2024}, is the eigenstructure of the covariance matrix. In the theoretical model of \gls{FAS}, densifying the aperture by adding many ports yields diminishing returns, as the limit performance is dictated by the number of dominant eigenvalues of the covariance matrix. In fact, for a very large number of ports in a finite aperture, the covariance becomes extremely rank-deficient. To get some insight, we plot in Fig. \ref{fig:eigCovariance} the eigenvalues of both $\bm{\Sigma}_{\widetilde{m}}^{(m)}$ and $\bm{\Sigma}_{\vec{y},\widetilde{m}}^{(m)}$, i.e., the eigenvalues of the resulting covariance and the wireless channel covariance. As observed, the number of relevant eigenvalues in both matrices is similar, which implies we are not losing degrees of freedom compared to the theoretical model of \gls{FAS}. In practice, this means there seemingly is nothing to lose in terms of performance, comparing the \gls{DMA}-based implementation with the original (theoretical) \gls{FAS} model, where the antenna can be freely moved along the aperture. However, we see that the magnitude of the eigenvalues decays faster in $\bm{\Sigma}_{\widetilde{m}}^{(m)}$. We analyze the impact of this on the \gls{FAS} performance in the next section.

%------------------------------------------------------------------------------------------------------------------
% PERFORMANCE OF DMA FAS
%------------------------------------------------------------------------------------------------------------------
\section{Performance of DMA-based FAS in FAMA}
\label{sec:Performance}
After mapping the circuital model into the standard \gls{FAS} communications model and analyzing the resulting end-to-end covariance structure, the last step is illustrating the impact of replacing an ideal fluid antenna---where the antenna element can be freely moved---by a \gls{DMA}-based implementation. To that end, consider the same scenario as in Section \ref{sec:FAMA_signal_model}, where a \gls{BS} with $M$ independent antennas serves $M$ users equipped with a \gls{DMA}-based \gls{FAS}. The received signal at user $m$ is then given by \eqref{eq:s}, where $u = 1,\dots \mathcal{U}$ indicates the codeword (configuration) employed by the user. 

The \gls{DMA} is the same as in Section \ref{sec:Covariance}, where the wireless channel is assumed Gaussian, i.e., $\vec{y}_{\text{ds},\widetilde{m}}^{(m)}\sim\mathcal{CN}_N(\vec{0},\bm{\Sigma}_\vec{y})$ for every $m,\widetilde{m}$, with $\bm{\Sigma}_\vec{y}$ as in \eqref{eq:SigmaYiso}. Assuming channel knowledge, each user selects the configuration that maximizes the \gls{SIR} (interference limited scenario), and hence the `port' (codeword) selection strategy is given by 
\begin{align}
    \widehat{u}_m = \mathop{\text{arg max}}_{u} \gamma_m, \quad \gamma_m = \frac{\left|\vec{g}_u^T\vec{y}_{\text{ds},m}^{(m)}\right|^2}{\sum_{\widetilde{m}\neq m} \left|\vec{g}_u^T\vec{y}_{\text{ds},\widetilde{m}}^{(m)}\right|^2}. \label{eq:SIR}
\end{align}

Note that channel knowledge is assumed as in the ideal \gls{FAS} related literature \cite{Wong2022}, since our objective here is to evaluate the impact of the \gls{DMA} design on the performance. Conceptually, channel estimation should not differ from the conventional one in \gls{FAS}, as ultimately the received signal at the different ports (configurations) is just a correlated random vector. However, this is not treated in detail here due to space constraints. 

From the port selection and the \gls{SIR} in \eqref{eq:SIR}, we focus on the evaluation of the outage probability, defined as $P(\gamma < \gamma_{\text{th}})$ for some threshold $\gamma_\text{th}$ by considering three different cases:
\begin{enumerate}
    \item \gls{DMA}-based \gls{FAS} using the whole codebook designed in Section \ref{sec:Covariance} and containing $\mathcal{U}=46$ different configurations.
    \item \gls{DMA}-based \gls{FAS} using a smaller codebook with $\mathcal{U}= 10$ configurations, which are obtained by sampling the previous codebook at 10 equally spaced angular directions. In principle, this yields less number of `ports' with less correlation, as the angular directions are more sparse. This baseline is useful to evaluate the impact of `densifying' the fluid antenna, analyzed in \cite[Section V]{Ramirez2024}.
    \item A theoretical (ideal) fluid antenna, as usually considered in the literature, where the equivalent channel is directly the wireless channel. Mathematically, this is done by removing the \gls{DMA} response $\vec{g}_u$ from \eqref{eq:s} and letting $\vec{y}_{\text{ds},\widetilde{m}}^{(m)}\sim\mathcal{CN}_U(\vec{0},\bm{\Sigma}_\text{iso})$ be the wireless channel to $\mathcal{U}=46$ equally spaced positions along the \gls{DMA} aperture. The covariance matrix $\bm{\Sigma}_\text{iso}$ is computed again as in \eqref{eq:SigmaYiso} but using the newly defined positions. This baseline emulates the theoretical performance of a freely-moving antenna along a linear aperture, as envisioned in, e.g., \cite{Wong2022}. 
\end{enumerate}

\begin{figure}[t]
    \centering
    \includegraphics[clip, trim={0 0.5cm 0 1cm}, width=1\linewidth]{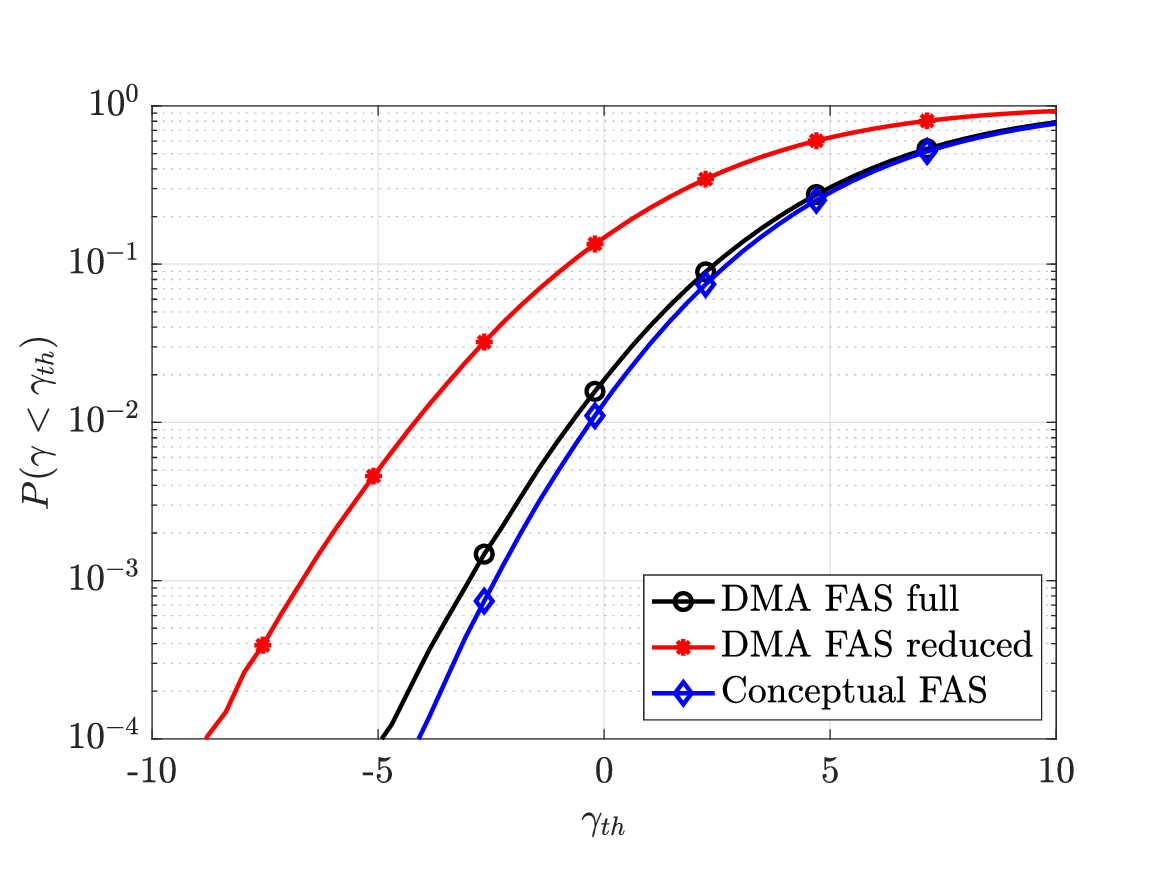}
    \caption{Outage probability achieved by a \gls{DMA}-based \gls{FAS} with a dense ($\mathcal{U}= 46$) and a reduced ($\mathcal{U}= 10$) codebook compared with a conceptual fluid antenna. Isotropic propagation is considered and the number of users is $M = 3$.}
    \label{fig:OP}
\end{figure}

The outage probability for the three schemes above is plotted in Fig. \ref{fig:OP} for $M = 3$. As expected from Fig. \ref{fig:eigCovariance}, the performance attained by the \gls{DMA}-based \gls{FAS} is close to that of an ideal \gls{FAS}, while reducing the codebook removes most of the gains as the granularity in the `port' selection is severely limited. That is, the gain brought by oversampling (by having enough `ports' or configurations) in \gls{FAS} essentially vanishes.

\begin{remark}
    Importantly, the results in Fig. \ref{fig:OP} illustrate that an electronically reconfigurable antenna, based on simple hardware designs, can achieve the envisioned performance of ideal \gls{FAS}, where antennas are physically moved. Note also that the codebook and the antenna designs can be further optimized to match desired correlation structures (as done in \cite{Zhang2024} for Jakes' correlation) or to emulate different correlation structures that would potentially yield improved performance. This emphasizes the importance of the covariance design open problem pointed out in Remark \ref{remark:Correlation}. 
\end{remark}

%------------------------------------------------------------------------------------------------------------------
% CONCLUSIONS
%------------------------------------------------------------------------------------------------------------------
\section{Conclusions}
\label{sec:Conclusions}
We have proposed and validated a comprehensive communications model for metasurface-based fluid antennas, demonstrating the potential of \glspl{DMA} to serve as practical and feasible embodiments of the fluid antenna concept. The model proposed here is not only capable of accurately characterizing specific designs, but also serves as a flexible mathematical tool for analyzing and modeling metasurface-based \glspl{FAS}. 

Despite the popularity of \glspl{FAS} in the wireless community, most of the works on performance analysis still build upon the idealized signal model where an antenna element can be freely moved along a given aperture with sub-wavelength resolution. However, such \gls{FAS} concept is far from reality, and electronically reconfigurable \glspl{FAS} arise as a more feasible alternative. Without the need to physically move the antenna elements, the proposed \gls{DMA}-based implementation provides a means to quickly switch (electronically) between different designed \glspl{FAS} configurations (i.e., \emph{virtual ports}), emulating the original (idealized) \glspl{FAS} concept in a practical manner. This could indeed open the door to realize the extremely promising fast-FAMA concept, with massive multiplexing capabilities, thus far deemed impractical due to the necessary extremely fast switching.

Our model can significantly simplify performance analysis and system design for these types of implementations, providing wireless researchers with a refined \gls{FAS} signal model that can be used for analysis using conventional approaches. Very importantly, the newly proposed signal model accounts for electromagnetic effects usually ignored, incorporating reflection coefficients, radiation pattern characteristics, or radiation efficiency in a natural way from a signal processing viewpoint.
Besides, the model can be used to analyze generic designs with arbitrary load admittances without relying on time-consuming full-wave simulations or intensive (costly) measurements, and only needs calibration if a specific radiating element is being characterized. Note also that, although we have stick to $2.4$ GHz as operating frequency, the model is valid for any frequency band including the promising FR3 band, and it is also valid for multiple \gls{RF} chains designs. 

This paper also unveils new challenges and open problems in \glspl{FAS}, such as codebook design and covariance analysis. Focusing on the latter, we show that the propagation environment is not the only factor impacting the end-to-end correlation, and that the resulting covariance can be tuned by changing the \gls{FAS} radiation properties. Hence, in principle, one could optimize \gls{FAS} performance by suitably designing different DMA-based configurations (codebook) that lead to desired structures of the effective correlation (covariance) matrix.

%------------------------------------------------------------------------------------------------------------------
% APPENDIX
%------------------------------------------------------------------------------------------------------------------
\appendices

\section{Channel modeling}
\label{app:ChannelModel}
\begin{figure}[t]
    \centering
    \includegraphics[width=0.5\linewidth]{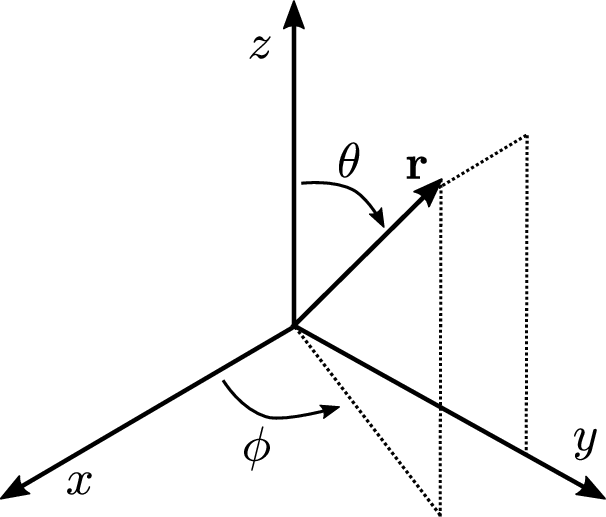}
    \caption{Azimuth ($\phi$) and polar ($\theta$) angles definition.}
    \label{fig:coordinates}
\end{figure}

The mutual admittance $\mat{Y}_\text{ds}$ representing the wireless channel can be written as $\mat{Y}_\text{ds} = \begin{pmatrix}\vec{y}_1  & \vec{y}_2 & \dots &\vec{y}_M\end{pmatrix}^T$, where $\vec{y}_m\in\mathbb{C}^{N\times 1}$ is the channel between device $m$ and the \gls{DMA}. A commonly accepted model assumes that $\vec{y}_m$ is given as the superposition of $P$ independent plane waves by (the dependence on $m$ is dropped)
\begin{equation}
    \vec{y} = \sum_{p=1}^P \frac{\alpha_p}{P}\sin(\theta_p)\sin(\vartheta_p)\vec{a}(\theta_p,\phi_p), \label{eq:GeneralChannelModel}
\end{equation}
where $\alpha_p\in\mathbb{C}$ is the complex amplitude of the $p$-th plane wave (capturing also polarization losses), $\theta_p$ and $\vartheta_p$ are the arrival and departure polar angles (see Fig. \ref{fig:coordinates}), $\phi_p$ is the arrival azimuth, and $\vec{a}(\theta_p,\phi_p)$ is the steering vector given by
\begin{equation}
    \vec{a}(\theta,\phi) = \begin{pmatrix}e^{i\vec{k}(\theta,\phi)^T\vec{r}_1} & e^{i\vec{k}(\theta,\phi)^T\vec{r}_2} & \dots & e^{i\vec{k}(\theta,\phi)^T\vec{r}_N}\end{pmatrix}^T
\end{equation}
with $\vec{k}(\theta,\phi) = k_0\begin{pmatrix}\sin(\theta)\cos(\phi) & \sin(\theta)\sin(\phi) & \cos(\theta)\end{pmatrix}^T$. Note that the $\sin(\cdot)$ terms in \eqref{eq:GeneralChannelModel} represent the normalized radiation pattern of the transmitting and receiving dipoles.

The parameters of the channel can be obtained either by measurements, simulations, or by assuming a statistical distribution. In the last case, we can particularize for the widely-used Rayleigh fading, in which $P\rightarrow \infty$ and $\alpha_p \sim\mathcal{CN}(0,\sigma_\alpha^2)$. Under these conditions, $\vec{y}$ is a multivariate Gaussian vector with zero mean and covariance matrix
\begin{equation}
    \bm{\Sigma}_\vec{y} = \sigma_\alpha^2\mathbb{E}_{\theta,\phi}\left[\sin(\theta)^2\vec{a}(\theta,\phi)\vec{a}(\theta,\phi)^H\right]\mathbb{E}_\vartheta\left[\sin(\vartheta)^2\right], \label{eq:CovExpectation}
\end{equation}
which is a function of the angular distribution and, consequently, the propagation environment. Isotropic scattering is usually assumed in \gls{MIMO} and \gls{FAS} literature, characterized by the \glspl{PDF}
\begin{align}
    f(\theta,\phi) =& \frac{\sin(\theta)}{2\pi}, & &0\leq\theta,\phi < \pi, \label{eq:PDF_theta_phi}\\
    f(\vartheta) =& \frac{\sin(\vartheta)}{2}, & &0\leq\vartheta < \pi,
\end{align}
leading to (see \cite[App. C]{Williams2022})
\begin{equation}
    (\bm{\Sigma}_\vec{y})_{n,n'} = -\frac{8\pi\sigma_\alpha^2}{3k_0}\imag{G_a(\vec{r}_n,\vec{r}_{n'})}. \label{eq:SigmaYiso}
\end{equation}

As important particular cases, removing the terms $\sin(\theta)$ and $\sin(\vartheta)$ from \eqref{eq:CovExpectation}---equivalently, replacing the dipoles by isotropic antennas---renders the classical $\text{sinc}(\cdot)$ correlation \cite[Eq. 7]{Ramirez2024}, while additionally enforcing $\theta = \pi/2$ (2D isotropic scattering) gives rise to Jakes' correlation function \cite[Eq. 8]{Ramirez2024}. Finally, note that the previous model does not depend on the dipole length $l_\text{m}$. However, one can weight $\vec{y}_\text{m}$ by $l_\text{m}^2$ to account for this factor. 

%------------------------------------------------------------------------------------------------------------------
% References
%------------------------------------------------------------------------------------------------------------------
%\vspace{10mm}

\bibliographystyle{IEEEtran}
\bibliography{references.bib}

\begin{thebibliography}{10}
\providecommand{\url}[1]{#1}
\csname url@samestyle\endcsname
\providecommand{\newblock}{\relax}
\providecommand{\bibinfo}[2]{#2}
\providecommand{\BIBentrySTDinterwordspacing}{\spaceskip=0pt\relax}
\providecommand{\BIBentryALTinterwordstretchfactor}{4}
\providecommand{\BIBentryALTinterwordspacing}{\spaceskip=\fontdimen2\font plus
\BIBentryALTinterwordstretchfactor\fontdimen3\font minus \fontdimen4\font\relax}
\providecommand{\BIBforeignlanguage}[2]{{%
\expandafter\ifx\csname l@#1\endcsname\relax
\typeout{** WARNING: IEEEtran.bst: No hyphenation pattern has been}%
\typeout{** loaded for the language `#1'. Using the pattern for}%
\typeout{** the default language instead.}%
\else
\language=\csname l@#1\endcsname
\fi
#2}}
\providecommand{\BIBdecl}{\relax}
\BIBdecl

\bibitem{Wong2022_BruceLee}
K.-K. Wong, K.-F. Tong, Y.~Shen, Y.~Chen, and Y.~Zhang, ``Bruce lee-inspired fluid antenna system: Six research topics and the potentials for {6G},'' \emph{Front. Comms. Net.}, vol.~3, 2022.

\bibitem{Wong2022}
K.-K. Wong, D.~Morales-Jimenez, K.-F. Tong, and C.-B. Chae, ``Slow fluid antenna multiple access,'' \emph{IEEE Trans. Commun.}, vol.~71, no.~5, pp. 2831--2846, 2023.

\bibitem{Ramirez2024}
P.~Ramírez-Espinosa, D.~Morales-Jimenez, and K.-K. Wong, ``A new spatial block-correlation model for fluid antenna systems,'' \emph{IEEE Trans. Wireless Commun.}, vol.~23, no.~11, pp. 15\,829--15\,843, 2024.

\bibitem{Zhu2024}
L.~Zhu, W.~Ma, and R.~Zhang, ``Modeling and performance analysis for movable antenna enabled wireless communications,'' \emph{IEEE Trans. Wireless Commun.}, vol.~23, no.~6, pp. 6234--6250, 2024.

\bibitem{New2024}
W.~K. New, K.-K. Wong, H.~Xu, K.-F. Tong, and C.-B. Chae, ``Fluid antenna system: New insights on outage probability and diversity gain,'' \emph{IEEE Trans. Wireless Commun.}, vol.~23, no.~1, pp. 128--140, 2024.

\bibitem{Khammassi2023}
M.~Khammassi, A.~Kammoun, and M.-S. Alouini, ``A new analytical approximation of the fluid antenna system channel,'' \emph{IEEE Trans. Wireless Commun.}, vol.~22, no.~12, pp. 8843--8858, 2023.

\bibitem{Zhu2024_movable}
L.~Zhu, W.~Ma, B.~Ning, and R.~Zhang, ``Movable-antenna enhanced multiuser communication via antenna position optimization,'' \emph{IEEE Trans. Wireless Commun.}, vol.~23, no.~7, pp. 7214--7229, 2024.

\bibitem{Zhu2024_magazine}
L.~Zhu, W.~Ma, and R.~Zhang, ``Movable antennas for wireless communication: Opportunities and challenges,'' \emph{IEEE Commun. Mag.}, vol.~62, no.~6, pp. 114--120, 2024.

\bibitem{Dong2024}
Z.~Dong \emph{et~al.}, ``Movable antenna for wireless communications: Prototyping and experimental results,'' \emph{arXiv e-prints, arXiv:2408.08588 [cs.IT]}, 2024.

\bibitem{Naqvi2019}
A.~H. Naqvi and S.~Lim, ``Fluidically beam-steering metasurfaced antenna,'' in \emph{2019 IEEE Int. Symp. Antennas Propag. USNC-URSI Radio Science Meeting}, 2019, pp. 695--696.

\bibitem{Huang2021}
Y.~Huang, L.~Xing, C.~Song, S.~Wang, and F.~Elhouni, ``Liquid antennas: Past, present and future,'' \emph{IEEE Open J. Antennas Propag.}, vol.~2, pp. 473--487, 2021.

\bibitem{Wang2025_arxiv}
R.~Wang \emph{et~al.}, ``Electromagnetically reconfigurable fluid antenna system for wireless communications: Design, modeling, algorithm, fabrication, and experiment,'' \emph{arXiv e-prints, arXiv:2502.19643 [eess.SP]}, 2025.

\bibitem{Shen2024}
Y.~Shen \emph{et~al.}, ``Design and implementation of mmwave surface wave enabled fluid antennas and experimental results for fluid antenna multiple access,'' \emph{arXiv e-prints, arXiv:2405.09663 [eess]}, 2024.

\bibitem{Liu2025}
B.~Liu, K.-F. Tong, K.-K. Wong, C.-B. Chae, and H.~Wong, ``Programmable meta-fluid antenna for spatial multiplexing in fast fluctuating radio channels,'' \emph{Opt. Express}, vol.~33, no.~13, pp. 28\,898--28\,915, Jun 2025.

\bibitem{Zhang2024}
J.~Zhang \emph{et~al.}, ``A novel pixel-based reconfigurable antenna applied in fluid antenna systems with high switching speed,'' \emph{IEEE Open J. Antennas Propag.}, vol.~6, no.~1, pp. 212--228, 2025.

\bibitem{Schlezinger2021}
N.~Shlezinger, G.~C. Alexandropoulos, M.~F. Imani, Y.~C. Eldar, and D.~R. Smith, ``Dynamic metasurface antennas for {6G} extreme massive {MIMO} communications,'' \emph{IEEE Wireless Commun.}, vol.~28, no.~2, pp. 106--113, 2021.

\bibitem{Carlson2024}
J.~Carlson, M.~R. Castellanos, and R.~W. Heath, ``Hierarchical codebook design with dynamic metasurface antennas for energy-efficient arrays,'' \emph{IEEE Trans. Wireless Commun.}, vol.~23, no.~10, pp. 14\,790--14\,804, 2024.

\bibitem{Williams2022}
R.~J. Williams, P.~Ramírez-Espinosa, J.~Yuan, and E.~de~Carvalho, ``Electromagnetic based communication model for dynamic metasurface antennas,'' \emph{IEEE Trans. Wireless Commun.}, vol.~21, no.~10, pp. 8616--8630, 2022.

\bibitem{You2023}
L.~You, J.~Xu, G.~C. Alexandropoulos, J.~Wang, W.~Wang, and X.~Gao, ``Energy efficiency maximization of massive {MIMO} communications with dynamic metasurface antennas,'' \emph{IEEE Trans. Wireless Commun.}, vol.~22, no.~1, pp. 393--407, 2023.

\bibitem{heath2025tri}
R.~W. Heath \emph{et~al.}, ``The tri-hybrid {MIMO} architecture,'' \emph{arXiv preprint arXiv:2505.21971}, 2025.

\bibitem{Ramirez2025}
P.~Ramírez-Espinosa, D.~Morales-Jiménez, and B.~Soret, ``Energy efficiency of {DMAs} vs. conventional {MIMO}: a sensitivity analysis,'' \emph{arXiv e-prints, arXiv:2506:09181 [eess.SP]}, 2025.

\bibitem{boyarsky2021_sciReports}
M.~Boyarsky, T.~Sleasman, M.~F. Imani, J.~N. Gollub, and D.~R. Smith, ``Electronically steered metasurface antenna,'' \emph{Scientific reports}, vol.~11, no.~1, p. 4693, 2021.

\bibitem{Sleasman2016}
T.~Sleasman \emph{et~al.}, ``Waveguide-fed tunable metamaterial element for dynamic apertures,'' \emph{IEEE Antennas Wireless Propag. Lett.}, vol.~15, pp. 606--609, 2016.

\bibitem{Yoo2023_AWPL}
I.~Yoo, D.~R. Smith, and M.~Boyarsky, ``Experimental characterization of a waveguide-fed varactor-tuned metamaterial element using the coupled dipole framework,'' \emph{IEEE Antennas Wireless Propag. Lett.}, vol.~22, no.~2, pp. 387--391, 2023.

\bibitem{Li2021_TAP}
S.~Li, F.~Xu, X.~Wan, T.~J. Cui, and Y.-Q. Jin, ``Programmable metasurface based on substrate-integrated waveguide for compact dynamic-pattern antenna,'' \emph{IEEE Trans. Antennas Propag.}, vol.~69, no.~5, pp. 2958--2962, 2021.

\bibitem{Jabbar2024}
A.~Jabbar \emph{et~al.}, ``60 {GHz} programmable dynamic metasurface antenna ({DMA}) for next-generation communication, sensing, and imaging applications: From concept to prototype,'' \emph{IEEE Open J. Antennas Propag.}, vol.~5, no.~3, pp. 705--726, 2024.

\bibitem{Wong2022_EL}
K.~K. Wong, K.~F. Tong, Y.~Chen, and Y.~Zhang, ``Closed-form expressions for spatial correlation parameters for performance analysis of fluid antenna systems,'' \emph{Electronics Lett.}, vol.~58, no.~11, pp. 454--457, 2022.

\bibitem{Smith2017}
D.~R. Smith, O.~Yurduseven, L.~P. Mancera, P.~Bowen, and N.~B. Kundtz, ``Analysis of a waveguide-fed metasurface antenna,'' \emph{Phys. Rev. Applied}, vol.~8, p. 054048, Nov 2017.

\bibitem{Johnson2014}
M.~Johnson, P.~Bowen, N.~Kundtz, and A.~Bily, ``Discrete-dipole approximation model for control and optimization of a holographic metamaterial antenna,'' \emph{Appl. Opt.}, vol.~53, no.~25, pp. 5791--5799, 2014.

\bibitem{Rogers1986}
P.~Rogers, ``Application of the minimum scattering antenna theory to mismatched antennas,'' \emph{IEEE Trans. Antennas Propag.}, vol.~34, no.~10, pp. 1223--1228, 1986.

\bibitem{Li1995}
L.~Li, P.~Kooi, M.~Leong, T.~Yeo, and S.~Ho, ``\BIBforeignlanguage{en}{On the eigenfunction expansion of electromagnetic dyadic {Green}'s functions in rectangular cavities and waveguides},'' \emph{\BIBforeignlanguage{en}{IEEE Trans. Microw. Theory Techn.}}, vol.~43, no.~3, pp. 700--702, Mar. 1995.

\bibitem{Wong2024}
K.-K. Wong, C.-B. Chae, and K.-F. Tong, ``Compact ultra massive antenna array: A simple open-loop massive connectivity scheme,'' \emph{IEEE Trans. Wireless Commun.}, vol.~23, no.~6, pp. 6279--6294, 2024.

\bibitem{Pozar2012}
D.~M. Pozar, \emph{Microwave engineering}, 4th~ed.\hskip 1em plus 0.5em minus 0.4em\relax Hoboken, NJ: Wiley, 2012.

\bibitem{Nossek2024}
J.~A. Nossek, D.~Semmler, M.~Joham, and W.~Utschick, ``Physically consistent modeling of wireless links with reconfigurable intelligent surfaces using multiport network analysis,'' \emph{IEEE Wireless Commun. Lett.}, vol.~13, no.~8, pp. 2240--2244, 2024.

\bibitem{yan2004simulation}
L.~Yan, W.~Hong, G.~Hua, J.~Chen, K.~Wu, and T.~J. Cui, ``Simulation and experiment on {SIW} slot array antennas,'' \emph{IEEE Microw. Wireless Compon. Lett.}, vol.~14, no.~9, pp. 446--448, 2004.

\bibitem{Pin_diode_model}
{Skyworks Solutions}, ``{SMP1345 Series Datasheet},'' \url{https://www.skyworksinc.com/Products/Diodes/SMP1345-Series}, accessed: 2025-07-10.

\end{thebibliography}

\end{document}